\documentclass{article}

\usepackage{jcappub} 
\usepackage{ulem}
\usepackage[utf8]{inputenc}
\usepackage{appendix}
\usepackage{ulem}

\title{Prospects for High-Elevation Radio Detection of $>$100 PeV Tau Neutrinos} 

\author[a,b,c]{Stephanie Wissel}\emailAdd{wissel@psu.edu}
\author[d,e]{Andr\'es Romero-Wolf}
\author[f]{Harm Schoorlemmer}
\author[g,h]{Washington R. Carvalho Jr.}
\author[h]{Jaime Alvarez-Mu\~niz}
\author[h]{Enrique Zas}
\author[i,j]{Austin Cummings}
\author[k]{Cosmin Deaconu}
\author[k,l]{Kaeli Hughes}
\author[k,l,m]{Andrew Ludwig}
\author[k]{Joalda Morancy}
\author[k]{Eric Oberla}
\author[a]{Caroline Paciaroni}
\author[n]{Steven Prohira}
\author[k,l]{Dan Southall}
\author[a]{Max Stapel-Kalat}
\author[m]{Ben Strutt}
\author[a]{Mercedes Vasquez}
\author[k,l,o]{Abigail Vieregg}

\affiliation[a]{Physics Department, California Polytechnic State University, San Luis Obispo, CA, 93407, USA}
\affiliation[b]{Department of Physics, Pennsylvania State University, State College, PA, 16802, USA}
\affiliation[c]{Department of Astronomy \& Astrophysics, Pennsylvania State University, State College, PA, 16802, USA}
\affiliation[d]{Jet Propulsion Laboratory, California Institute of Technology, Pasadena, CA 91109, USA}
\affiliation[e]{Division of Physics, Mathematics, and Astronomy, California Institute of Technology, Pasadena, CA 91125, USA}
\affiliation[f]{Max-Planck-Institut f\"ur Kernphysik, 69117, Heidelberg, Germany}
\affiliation[g]{Departamento de F\'\i sica,
Universidade de S\~ao Paulo, S\~ao Paulo, Brazil}
\affiliation[h]{Instituto Galego de F\'\i sica de Altas Enerx\'\i as (IGFAE),
Universidade de Santiago de Compostela, 15782 Santiago
de Compostela, Spain}
\affiliation[i]{Gran Sasso Science Institute, L’Aquila, Italy}
\affiliation[j]{INFN-Laboratori Nazionali Gran Sasso, Assergi (AQ), Italy}
\affiliation[k]{Kavli Institute for Cosmological Physics, University of Chicago, Chicago, IL 60637 USA}
\affiliation[l]{Department of Physics, University of Chicago, Chicago, IL 60637 USA}
\affiliation[m]{Department of Physics and Astronomy, University of California, Los Angeles, Los Angeles, CA 90095 USA}
\affiliation[n]{Department of Physics, Center for Cosmology and AstroParticle Physics, Ohio State University, Columbus, OH 43210 USA}
\affiliation[o]{Enrico Fermi Institute, University of Chicago, Chicago, IL 60637 USA}

\begin{document}
\abstract{
     Tau neutrinos are expected to comprise roughly one third of both the astrophysical and cosmogenic neutrino flux, but currently the flavor ratio is poorly constrained and the expected flux at energies above $10^{17}$~eV is low. We present a detector concept aimed at measuring the diffuse flux of tau neutrinos in this energy range via a high-elevation mountaintop detector using the radio technique. The detector searches for radio signals from upgoing air showers generated by Earth-skimming tau neutrinos. Signals from several antennas in a compact array are coherently summed at the trigger level, permitting not only directional masking of anthropogenic backgrounds, but also a low trigger threshold. This design takes advantage of both the large viewing area available at high-elevation sites and the nearly full duty cycle available to radio instruments. We present trade studies that consider the station elevation, frequency band, number of antennas in the array, and the trigger threshold to develop a highly efficient station design. Such a mountaintop detector can achieve a factor of ten improvement in acceptance over existing instruments with 100 independent stations. With 1000 stations and three years of observation, it can achieve a sensitivity to an integrated $\mathcal{E}^{-2}$ flux of $<10^{-9}$~GeV cm$^{-2}$ sr$^{-1}$ s$^{-1}$, in the range of the expected flux of all-flavor cosmogenic neutrinos assuming a pure iron cosmic-ray composition.
}

\maketitle
\section{Introduction}

Observations of cosmic neutrinos with energies $>10^{17}$~eV could provide information about high-energy astrophysical sources~\cite{Astro2020_nuastro} as well as the fundamental physics of electroweak interactions~\cite{Astro2020_nufundamental}. These neutrinos can be either produced at the source~\cite{Waxman_Bahcall_1999} or during the propagation of ultra-high energy cosmic rays interacting with cosmic background photons ~\cite{Berezinsky_Zatsepin_1969}. 

Recently, IceCube detected an astrophysical neutrino flux with energies between a few tens of TeV to $>$~PeV~\cite{Aartsen2014}. Although neutrinos of astrophysical origin are expected to be produced in predominantly $\nu_e$ and $\nu_\mu$ flavors, neutrino mixing over cosmological distances is expected to result in a flux at Earth with ratio $\nu_e:\nu_\mu:\nu_\tau=1:1:1$~\cite{Beacom_2003}. Detection of $\nu_\tau$'s would serve as additional proof that the neutrino flux is astrophysical since the electron and muon neutrinos produced in the atmosphere do not propagate for sufficiently long distances to produce significant mixing. Tau neutrino searches in the $\sim$1~km${^3}$ instrumented volume of the IceCube detector are challenging because $\nu_\tau$ cascade events are comparable in morphology to $\nu_e$ cascades. Although recent searches for a $\nu_\tau$ cascade followed by a $\tau$-decay-induced cascade are promising, they have a signal rate that is comparable to background~\cite{Stachurska_ICRC_2019, Xu_ICRC_2019}.
Measurements of the neutrino flux at energies $>10^{16}$~eV would serve to constrain a spectral cutoff~\cite{Haack_ICRC_2017} and further characterize sources of the IceCube flux. 

The Pierre Auger Observatory (hereafter Auger) has placed the most stringent limits on the single-flavor $\nu_\tau$ flux at energies $>10^{17}$~eV~\cite{Auger2019}. Although the sensitivity of the $\sim$3000~km$^2$ surface detector of Auger is comparable to IceCube, Auger is mostly sensitive to $\nu_\tau$'s while the high-energy limits placed by IceCube have limited neutrino-flavor identification capability. ANITA has placed the most stringent limits at energies $>10^{19}$~eV~\cite{Gorham2019} by searching for radio impulsive transients produced by neutrino interactions in the Antarctic ice. Recently, ANITA has reported the detection of two events consistent with upgoing air showers~\cite{Gorham2016, Gorham2018}. Although the phenomenology is consistent with an EeV-scale $\nu_\tau$ interaction in the Earth producing a $\tau$ lepton that exits the Earth and decays in the atmosphere, the sensitivity estimates are strongly discrepant with the diffuse flux limits of Auger and IceCube~\cite{Romero-Wolf2019,IceCubeANITAFollowup2020}.  

A new detector is clearly necessary to achieve the detection of the $\nu_\tau$ component of the flux and extend sensitivity to higher energies. An extension to the IceCube detector is being proposed to increase the volume of the array by nearly an order of magnitude with a lower density of detector strings as part of the IceCube-Gen2 program~\cite{IceCubeGen2}. This detector is expected to be sensitive to the neutrino flux for energies greater than $\sim 10^{16}$~eV and is capable of discriminating between $\nu_\mu$ and cascade-like events, providing constraints in flavor ratio changes. The ability to identify $\nu_\tau$ events with a low background remains to be demonstrated. Detectors such as those used in the ARA~\cite{ARA2015} and ARIANNA~\cite{ARIANNA2015} experiments use radio antennas several hundred meters in the ice and near the surface, respectively, to extend the energy sensitivity above $10^{17}$~eV. These detectors benefit from being in a quiet radio environment and could serve as a high-energy extension to IceCube. Other techniques~\cite{ProhiraRadar2020} and experimental designs~\cite{RNO2019, ARIANNA_ASR_2019} searching for neutrino interactions in ice are also being explored. At energies $>10^{17}$~eV, the GRAND array~\cite{GRAND_whitepaper} is being proposed to detect $\tau$-lepton air showers of $\nu_\tau$ origin. The detector consists of an array of 200,000 antennas spread over an area of 200,000~km$^2$ to observe the geomagnetic radio emission of grazing $\tau$-lepton air showers. 

We present a detector concept that makes use of beamformer arrays of antennas at high-elevation sites.  A detector at high elevation  with a clear view of the horizon has an acceptance that rises with detector height $h^{3/2}$~\cite{Motloch2014}, increasing the overall sensitivity per station significantly compared to antennas on the ground. Other proposed experiments like the optical experiments ASHRA-NTA~\cite{Sasaki2017} and TRINITY~\cite{Otte2018} and the radio experiment TAROGE~\cite{Nam_ICRC_2019} plan  to exploit the high-elevation concept; however, this study presumes the use of a beamformer radio array. The use of beamformer arrays is motivated by their improved trigger sensitivity~\cite{Romero-Wolf2015a, Vieregg2016}, and real-time rejection of backgrounds due to radio frequency interference (RFI) using coarse pointing at the trigger level. It is expected that RFI will be the limiting factor even in the most remote radio-quiet regions of Earth~\cite{Monroe2019}. Consisting of a modest sized beamformed array ($\sim$10 clustered antennas), each station is expected to be low cost and scalable such that stations can be installed in multiple sites around the globe. This would provide the benefit of a high acceptance global network of stations to build up full-sky coverage, while the use of the radio technique would allow continuous observations and nearly full duty cycles. Assuming $\nu_\tau$'s are discovered, pointing requirements can be met with antennas that trigger off the main cluster but are separated by long baselines as required by future investigations. 

This paper is organized as follows. In Section \ref{sec:detector} we present the detector concept and establish the figures of merit. In Section \ref{sec:acceptance} we describe the simulations done to arrive at two reference designs that demonstrates the detector's capabilities and strengths, both of which are presented in Section \ref{sec:refdesign}. In Section \ref{sec:design_vars} we present some of the design variations showing the flexibility of this concept to adapt to the characteristics of future potential observatory sites. In Section \ref{sec:discussion} we discuss the implications of the design and future work, and we conclude in Section \ref{sec:conclusion}. 
\section{Detector Concept}\label{sec:detector}

\begin{figure}[t]
\begin{center}
\includegraphics[height=0.29\textheight]{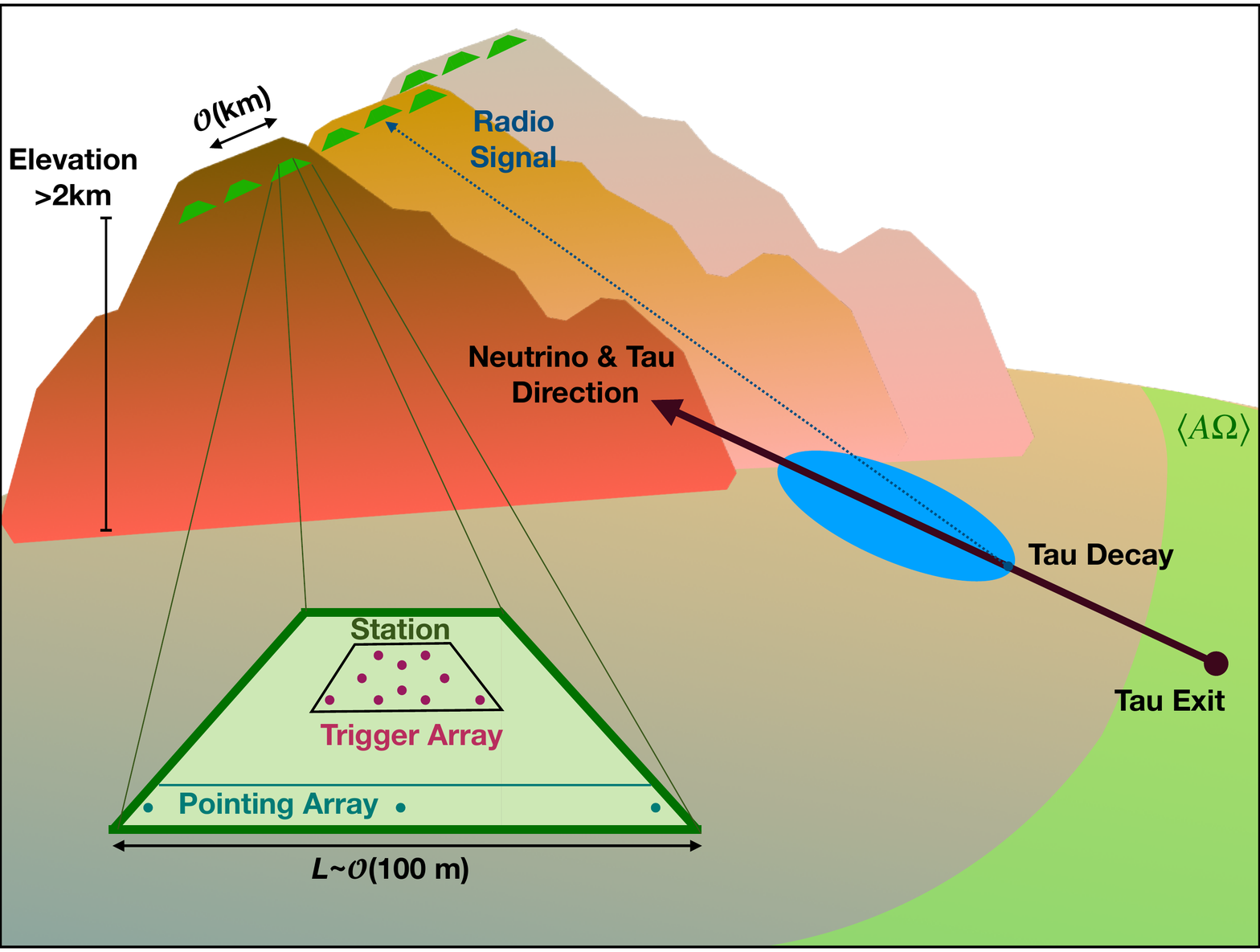}
\includegraphics[height=0.29\textheight]{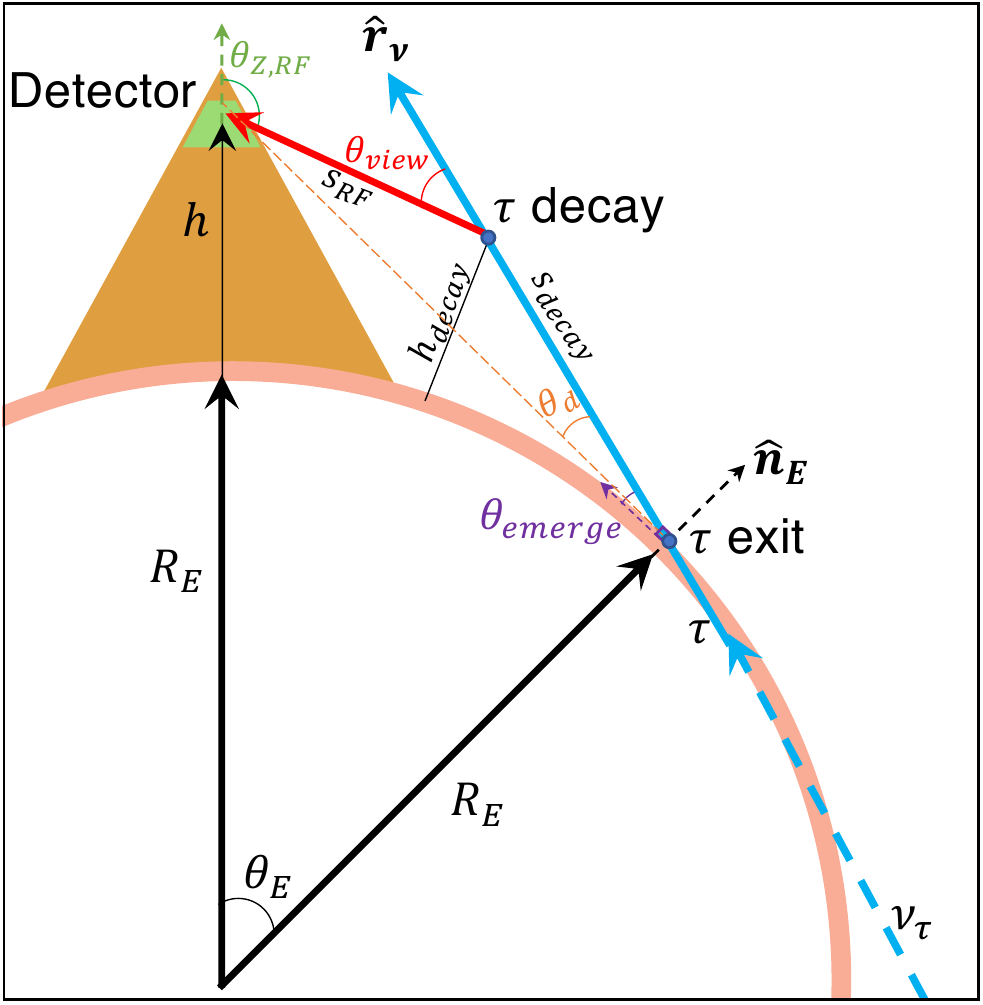}
\end{center}
\caption{ Observation concept for $\nu_\tau$'s from a high-elevation antenna array. A $\nu_\tau$ undergoes a charged-current interaction in the Earth producing a $\tau$ lepton. The $\tau$ lepton may then escape the solid layers of the Earth at the exit point at an Earth angle $\theta_E$ and exiting with a local emerging angle of $\theta_\mathrm{emerge}$ with respect to the horizontal. The $\tau$ lepton can then decay in the atmosphere at an altitude $h_{decay}$ producing an up-going extensive air shower. The air shower interacts with the Earth's magnetic field to produce a radio impulse that propagates and can be detected by a mountaintop station at a zenith angle $\theta_{Z,RF}$. The radio signal is emitted at an angle $\theta_{view}$ relative to the $\tau$ direction. Each station at elevation $h$ consists of $\mathcal{O}(10)$ antennas, arranged in a compact trigger array and a pointing array with longer baselines.}
\label{fig:geom_concept}
\end{figure}
 The concept presented here uses a radio beamformer at a high elevation site with a clear view of the horizon. As shown in Figure~\ref{fig:geom_concept}, the high elevation site permits the detector to monitor a large area over which $\nu_\tau$'s may interact in the Earth. The neutrino interaction produces a $\tau$ lepton that can exit into the atmosphere, decay to produce an extensive air shower, which would subsequently emit a radio impulse that can be observed by a detector at elevation.  Recent analyses have shown that the impulsive signals expected from air shower events can be identified using polarization and/or directional reconstruction in experiments that trigger only on the radio-frequency signals generated by the showers~\cite{Hoover2010,Ardouin2011,ARIANNA_CRs,Monroe2019}. 

Mountaintop detectors balance a large single-station acceptance with the continuous duty cycles available to ground-based detectors. A single station placed at 3~km views a surface area of the Earth of $\sim120,000$~km$^2$. Covering a comparable area with detectors on the ground would be a complex problem. The area visible to a single station at balloon altitudes is significantly higher, roughly $1.4\times 10^6$~km$^2$ for an altitude of 37~km. However, a ground-based station at elevation, with nearly 100\% duty cycle, compensates for the month-long flight times available each year. Additionally, a mountaintop detector is more sensitive to lower energy air showers than a balloon-based detector, because it is closer to the shower origin.

Given the high efficiency of a mountaintop detector, it is worth considering a design that minimizes the total number of required channels to either constrain models or discover ultra-high energy neutrinos. Arrays of radio antennas can be used in different configurations at a given station to observe radio signals from particle showers depending on the goals of the detector. Some radio experiments~\cite{Nelles2014, Monroe2019, GRAND_whitepaper} use a large number of antennas packed in a small number of stations to observe the footprint of the air shower on the ground. Other experiments \cite{Gorham2009a, ARA2015, ARIANNA2015}, use a small number of antennas per station to search for emission from the whole shower generated by a neutrino interaction. 

For the purposes of this study, multiple independent stations will comprise the full high-elevation detector. The stations are considered independent if they view independent target volumes. Each station consists of a beamformer array of antennas, a pointing array of antennas, which are triggered and recorded by a central data acquistion system. The beamformer array includes a group of antennas that operate together to detect a shower via real-time phasing built into the trigger firmware. An antenna in the beamformer array is assumed to be dual-polarization (horizontal and vertical) and therefore has two radio-frequency receiving channels. 
We will consider the optimization between a few stations each composed of large beamforming arrays or a large number of stations composed of a small number of beamforming arrays.
In this study, we limit the total number of channels to $\mathcal{O}$(1000), comparable to the number of channels in existing detectors like Auger and IceCube. 

A station must be able to reconstruct the neutrino arrival direction and that requirement sets the minimum number of antennas. Even a small antenna array consisting of 3 to 4 antennas is capable of degree-scale reconstruction or better. Figure~\ref{fig:geom_concept} shows a possible layout of an array that includes both a compact cluster with antennas separated by tens of meters (the beamformer array) and a pointing array that places outrigger antennas at $\mathcal{O}(100)$~m to improve pointing resolution. Increasing the total number of antennas in a given beamformer array can both lower the trigger threshold and improve the pointing resolution.

The beamformer array can be implemented at the trigger level as a phased array. In a phased array, beams are formed digitally and in real-time by delaying and summing waveforms from multiple antennas to increase the amplitude signal-to-noise ratio by a factor of $\sqrt{N}$, where $N$ is the number of antennas included in the trigger~\cite{Franceschetti2005}. Full coverage of the required solid angle can be achieved by including multiple beams, corresponding to different time delays among antennas, in the trigger. By down-weighting the contribution of certain beams in the trigger, clustered sources of man-made radio frequency interference (RFI) can be discriminated against in real time, while retaining sensitivity to air shower pulses. The antennas are tightly clustered in the trigger array, because the phasing technique presumes that each antenna in the array views a similar portion of the shower and is at a similar distance from the shower axis. In the mountaintop geometry, the detector triggers on showers that are between $\sim$10~km and 80~km away with lower energy showers triggering at closer distances and therefore steeper angles below the horizon. Thus, the showers are sufficiently far away to assume that a tightly clustered triggered array samples the radio emission from similar portions of the shower.  

The arrival directions of the radio signal can be reconstructed by using the relative arrival times between antennas observing the same signal. The pointing resolution is roughly $\delta\theta \sim c/(L \, \Delta f \, {\rm SNR})$, where $c$ is the speed of light, $L$ is the largest separation distance between two antennas in the array, $\Delta f$ is the bandwidth of the impulse, and ${\rm SNR}$ is the voltage signal-to-noise ratio in the beam referenced to a thermal noise of $\sigma$~\cite{Romero-Wolf2015a}. A beamformer array with a bandwidth of 50~MHz and maximum baseline of 70~m can achieve a pointing resolution of $\sim1^{\circ}$ on a 5$\sigma$ pulse in the beam. The layout of the array will weight the contributions of different baselines to determine the central beam pattern and its sidelobes. The neutrino arrival direction can be estimated using the arrival direction of the radio signal and an estimate of the angle between the radio signal and the direction of the original neutrino. Given that the Cherenkov angle is typically $\sim1^{\circ}$ in air, the expected uncertainty on the arrival direction is expected to be at the degree scale.
  
  A compact array with degree-scale radio pointing resolution is sufficient for RFI rejection as well as detection and identification of $\tau$-lepton air showers. However, if a $\nu_\tau$ flux at energies $>10^{17}$~eV is discovered, the pointing can be improved by adding antennas with longer separations from the main cluster. This would enable sub-degree pointing resolution to improve the test for point source searches. 
  
 The relatively low resource demands needed for a small cluster of antennas means that the array is expandable to multiple sites. The advantage is that the sensitivity to a diffuse flux can be increased better than linearly with time as scientifically significant flux limits are produced. This is an important feature given that the minimal predictions for the cosmogenic neutrino fluxes~\cite{Kotera2010, Batista_2019,vanVliet2019} are two orders of magnitude below today's best limits~\cite{Gorham2019, Auger2019, IceCubeUHE2018}.

With these insights in mind, the detector figures of merit include the gain of the antennas, frequency of operation, number of antennas in the beamformer array, and detector elevation. Aside from the need to point in real time for RFI rejection, the number of antennas increases the gain of the receiver, which improves the sensitivity to lower energy air showers. However, this can be compared to using the same number of antennas to create more stations, which improves the overall acceptance of the detector. Similarly, increasing the elevation of the detector improves the acceptance for high-energy showers, while reducing it improves the sensitivity to lower-energy showers. These trades will be discussed in more detail in the sections that follow. 

\section{Acceptance Estimates}\label{sec:acceptance}

The sensitivity of a detector tuning in to a diffuse flux of $\nu_\tau$ particles depends on the integrated time, area, and solid angle for observing the events. In this section, we present the formalism used for estimating the sensitivity, provide some first-order estimates, and describe the simulations used for the estimating of the acceptance $\langle A \Omega\rangle(\mathcal{E}_{\nu})$ for a given $\nu_\tau$ energy $\mathcal{E}_{\nu}$. The estimates presented here follow the approach used for the diffuse $\nu_\tau$ flux limits with ANITA~\cite{Romero-Wolf2019}.

The differential acceptance $d\langle A\Omega\rangle$ for a neutrino of energy $\mathcal{E}_{\nu}$ crossing an area element $dA$ with normal vector $\mathbf{\hat n_{E}}$ for neutrino propagation direction $\mathbf{\hat r}_\nu$ (see Fig.~\ref{fig:geom_concept}) with differential solid angle $d\Omega_\nu$ is
\begin{equation}
d\langle A\Omega\rangle = dA\,d\Omega_\nu \, \mathbf{\hat n_E} \cdot \mathbf{\hat r}_\nu \, p_\mathrm{obs},
\end{equation}
where $p_\mathrm{obs}$ is the probability that the particle with trajectory crossing the area element $dA$ with direction $\mathbf{\hat r}_\nu$ will be detected. 

For $\tau$ leptons produced by $\nu_\tau$ interactions in Earth, we choose the reference area of the Earth. The total acceptance to a diffuse neutrino flux is 
\begin{equation}
\langle A\Omega\rangle(\mathcal{E}_{\nu}) = R_E^2\iint d\Omega_E \iint d\Omega_\nu \, \mathbf{\hat n_E} \cdot \mathbf{\hat r}_\nu \, p_\mathrm{obs}(\mathcal{E}_{\nu}, \mathbf{x}_E, \mathbf{\hat r}_\nu),
\label{eq:acceptance_simple}
\end{equation}
where $\mathbf{x}_E$ is a position on the surface of the Earth at which the $\tau$ lepton exits and $R^2_E d\Omega_E$ is the differential area element on the surface of the Earth and $R_{E}$ is the radius of the Earth. 

The probability of observation $p_\mathrm{obs}(\mathcal{E}_{\nu}, \mathbf{x}_E, \mathbf{\hat r}_\nu)$ depends on a number of effects. The first is the probability that a $\nu_\tau$ produces a $\tau$ lepton of energy $\mathcal{E}_\tau$ that exits into the atmosphere. We may express the probability density function as $p_\mathrm{exit}(\mathcal{E}_\tau|\mathcal{E}_{\nu}, \theta_\mathrm{emerge})$, where $\theta_\mathrm{emerge}$ is the angle between the particle trajectory and the horizontal at the exit point. 
This is obtained using the NuTauSim code~\cite{Alvarez-Muniz2018}. 

After exiting into the atmosphere and traveling a distance $s_\mathrm{decay}$, the $\tau$ lepton decays  with a probability density function $p_\mathrm{decay}(s_\mathrm{decay}|\mathcal{E}_\tau)=\exp{[-s_\mathrm{decay}/D(\mathcal{E}_\tau)]}/D(\mathcal{E}_\tau)$, where $D(\mathcal{E}_\tau)$ is determined from the $\tau$ lepton lifetime to be $D(\mathcal{E}_\tau)=4.9~\mathrm{km}~(\mathcal{E}_\tau/10^{17}~\mathrm{eV})$. 

The $\tau$ lepton decay modes and the fractional energy distributions of particles that will contribute to an extensive air shower are obtained using PYTHIA~\cite{Sjostrand2015}. The probability distribution for the extensive air shower energy $\mathcal{E}_\mathrm{EAS}$ that results from $\tau$ decays is $p_\mathrm{EAS}(\mathcal{E}_\mathrm{EAS}|\mathcal{E}_\tau)$. The probability distribution generally depends on the geometry and decay position of the emerging $\tau$ lepton to account for variations in atmospheric density.

An extensive air shower interacts with the Earth's magnetic field to produce an electromagnetic impulse at the location of the detector with peak value $E_\mathrm{pk}(\mathbf{x}_\mathrm{det})$ through the combined effects of geomagnetic radiation and Askaryan radiation~\cite{Alvarez-Muniz2012}. A shower of energy $\mathcal{E}_\mathrm{EAS}$ will have fluctuations in its longitudinal distribution function, particularly in the position of shower maximum ($\chi_\mathrm{max}$) resulting in a probability distribution function for the electromagnetic emission $p_\mathrm{pk}(E_\mathrm{pk}| \mathcal{E}_\mathrm{EAS}, s_\mathrm{decay}, \mathbf{\hat r}_\nu)$. The electric field generated depends on $s_\mathrm{decay}$ and $\mathbf{\hat r}_\nu$, because the detector can be in the near field or far field of the shower, because the radio signal from certain portions of the shower can be coherent, and because the shower development depends on the atmospheric profile. 

Finally, when the electric field $\mathbf{E}(\mathbf{x}_\mathrm{det})$ excites an antenna, it will trigger an event with probability $p_\mathrm{trig}(\mathbf{x}_\mathrm{det}|E_\mathrm{pk})$. The probability that the detector is triggered depends on the ambient noise properties, the system temperature of the detector, and the antenna gain. The detector model will be treated in Section~\ref{Section:Detector}.

Altogether we may write the probability of observation as a nested integral 
\begin{equation}
\begin{split}
\langle A\Omega\rangle
(\mathcal{E}_{\nu})
= 
& R_E^2 \iint d\Omega_E \iint d\Omega_{\nu} \ \hat{\mathbf{r}}_{\nu}\cdot \hat{\mathbf{n}}_E 
\int d\mathcal{E}_{\tau } 
\ p_{\mbox{\scriptsize exit}}(\mathcal{E}_{\tau}|\mathcal{E}_{\nu}, \theta_{\mbox{\scriptsize emerge}}) 
\int ds_\mathrm{decay} \ p_\mathrm{decay}(s_\mathrm{decay}|\mathcal{E}_{\tau}) 
\\
& 
\int d\mathcal{E}_\mathrm{EAS} \ p_\mathrm{EAS}(\mathcal{E}_\mathrm{EAS}|\mathcal{E}_{\tau}) 
\int dE_\mathrm{pk} \ p_\mathrm{pk}(E_\mathrm{pk}|\mathcal{E}_\mathrm{EAS}, s_\mathrm{decay}, \hat{\mathbf{r}}_{\nu}) \
p_\mathrm{trig}(\mathbf{x}_\mathrm{det}| E_\mathrm{pk}) \\
\end{split}
\label{eq:acceptance}
\end{equation}

\subsection{First-Order Acceptance Estimate}

As a starting point, we can gain insight into the maximum acceptance achievable with a beamformer array at elevation through a first-order estimate based on the geometry, the exit probabilities, and a generous estimate of the observation probability that assumes all events within an angle $\theta_\mathrm{cut}$ are detected, and that all events decay before reaching the detector. 

We begin by calculating the maximum acceptance $\langle A\Omega \rangle_g$ allowed by geometry of any detector at an elevation $h$. We assume that all $\tau$ leptons decay at the exit point and that $\theta_\mathrm{cut}$ sets the maximum angle $\theta_d$ over which the showers are detectable, where $\theta_d$ is the angle between the neutrino trajectory and a straight line to the detector (see Fig.~\ref{fig:geom_concept}).  Under such assumptions, the geometric acceptance becomes

\begin{equation}
\langle A\Omega \rangle_g
=  R_E^2 \iint d\Omega_E \iint d\Omega_{\nu} \ \hat{\mathbf{r}}_{\nu}\cdot \hat{\mathbf{n}}_E \ 
\Theta(\theta_\mathrm{cut}-\theta_\mathrm{d})
\label{eq:geom_acceptance}
\end{equation}
where $\Theta$ is the Heaviside step function. The integral has been approximated in analytical form in Ref.~\cite{Motloch2014} to be: 

\begin{equation}
\langle A\Omega \rangle_g \simeq 2\pi^2\sin^2\theta_{\mathrm{cut}}
\frac{[h(2R_E+h)]^{3/2} -h^2(3R_E+h)}{3(R_E+h)}.
\label{eqn:crude_acceptance}
\end{equation}

Note, however, that while this is a good approximation for observatories at stratospheric balloon or low-Earth orbit altitudes, it systematically underestimates the geometric acceptance at mountain-top elevations by down-weighting contributions to the acceptance at large $\theta_\mathrm{cut}$. See Appendix~\ref{app:geometric_acceptance} for more details.

\begin{figure}[!t]
\centerline
{\includegraphics[width=0.7\textwidth]{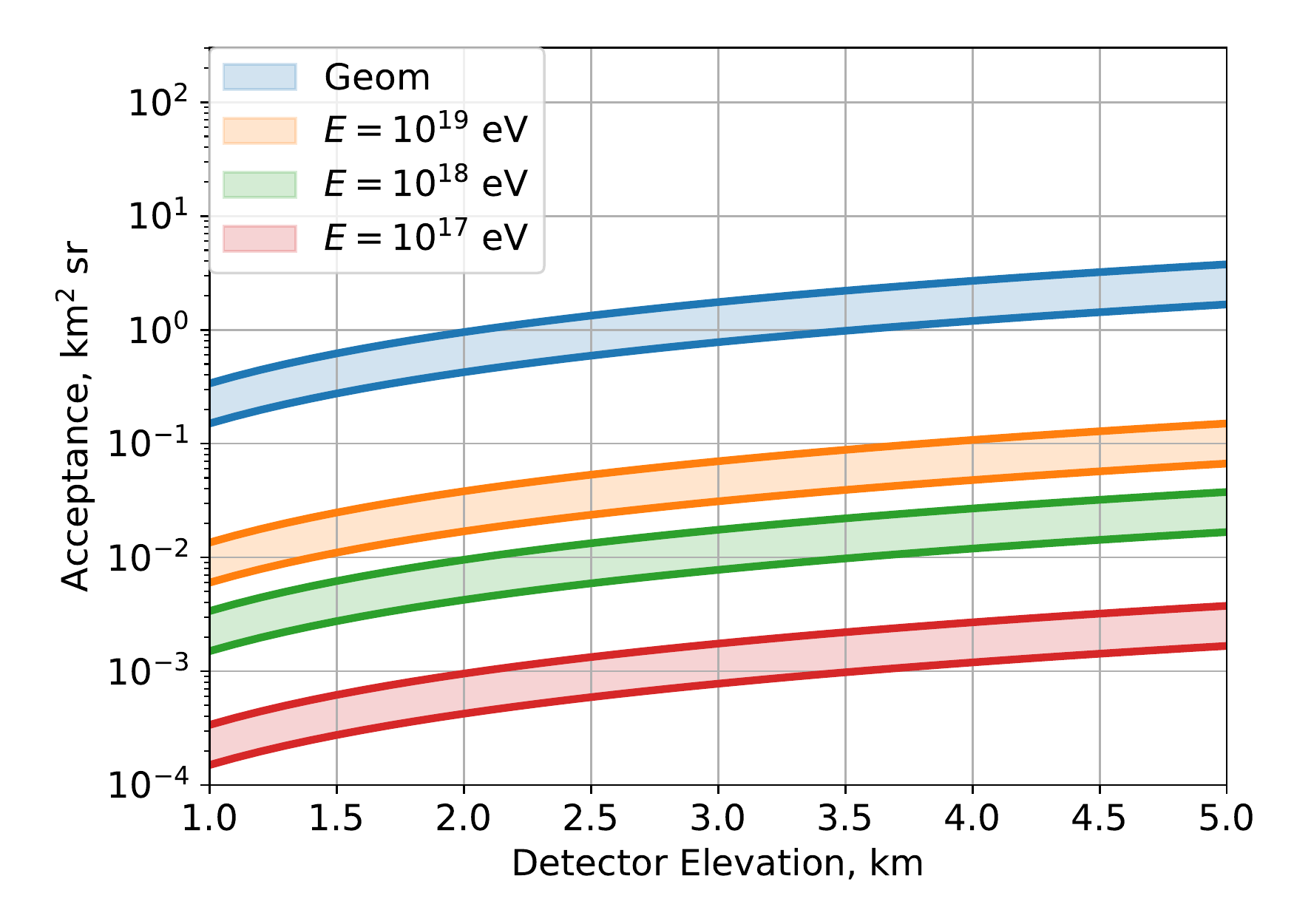}}
\caption{Order of magnitude estimates for the acceptance of a single high elevation station. The blue band shows the maximum geometric acceptance based on Eqn~\ref{eqn:crude_acceptance}, assuming a factor of $1/3$ to account for an expected 120$^\circ$ field of view. The acceptance at three neutrino energies $\mathcal{E}_{\nu}$ is estimated from peak $p_\mathrm{exit}$ values of Ref.~\cite{Alvarez-Muniz2018}. The width of each band illustrates the impact of the maximum view angle over which the detector can trigger $\theta_\mathrm{cut}$, assuming a range of $[1.0^\circ, 1.5^\circ]$.}
\label{fig:crude_estimate}
\end{figure}

 The acceptance range assuming values of $\theta_\mathrm{cut} \in [1.0^\circ, 1.5^\circ]$ is shown in Fig.~\ref{fig:crude_estimate} in the blue band. We assume an azimuthal field of view $\alpha=120^{\circ}$ . This is a practical approach so as not to require that all detectors be built on mountain peaks. Because the geometric aperture grows as $h^{3/2}$, the total acceptance increases by a factor of 2.5 going from 1~km to 2~km and by an overall factor of 7 going from 1~km to 4~km. 
 
 We can refine the acceptance estimate by including first-order approximations on the probability that a $\tau$ lepton will exit. Taking the peak values of the exit probability from Ref~\cite{Alvarez-Muniz2018}, we assume that exit probability is uniform in $\theta_\mathrm{emerge}$, but varies depending on the neutrino energy to arrive at the first-order estimates for three different energies shown in Fig.~\ref{fig:crude_estimate}. We can expect a detector comprising 100 stations at a 3~km elevation to achieve an acceptance at $10^{17}$~eV of at most 0.2~km$^2$sr. Due to the increased exit probability, the maximum acceptance grows to 7~km$^2$sr at $10^{19}$~eV.

\subsection{Monte Carlo Acceptance Estimates}\label{sec:firstorder}

To further refine the expected acceptance of a mountaintop detector to an isotropic  $\nu_\tau$ flux, we have evaluated Eqn.~\ref{eq:acceptance} using Monte Carlo integration techniques. This requires estimating several energy-dependent probabilities, including the probability that a $\tau$ lepton will exit the Earth, $p_{\mathrm{exit}}$, the probability that the $\tau$ lepton will decay before reaching the detector, $p_{\mathrm{decay}}$, the probability distribution for extensive air showers resulting from a given $\tau$ lepton decay, $p_{\mathrm{EAS}}$, and the probability that an air shower will trigger the detector, $p_{\mathrm{trig}}$. 

\subsubsection{Geometry}
The detector geometry is modeled as a single station at elevation $h$ viewing a spherical Earth with a mean Earth radius $R_E=6371$~km (see Figure~\ref{fig:geom_concept}).  To make the evaluation of the integral Eqn.~\ref{eq:acceptance} efficient, only the surface of the Earth in view of the detector at height $h$ where the $\tau$-lepton air shower may be observed has to be considered. This is implemented in the Monte Carlo by first sampling all $\theta_d$ and applying a cut on $\theta_d$ of $\theta_{cut}=5^{\circ}$. 
The latitudinal range of integration is $\theta_E\in[0,\theta_\mathrm{horizon}]$, where $\theta_E=0$ is the reference latitude of the detector (or Earth angle in Fig.~\ref{fig:geom_concept}) and $\theta_\mathrm{horizon}=\cos^{-1}(1/(1+h/R_E))$. No landforms are included in the simulations since these are site specific, but they will be added to future studies for specific candidate sites as was done in other studies of different radio instruments~\cite{Decoene2019}.

The simulation tracks the exit point and associated emergence angle, the position of the decay point, and the view angle with respect to the decay point. The distances from the exit and decay points to the detector, the altitude above sea level of the decay point, and the zenith angle of the radio emission at the detector are calculated assuming a spherical Earth geometry as depicted in Fig.~\ref{fig:geom_concept}. As in the first-order calculations, we assume an azimuthal field of view $\alpha$ of $120^{\circ}$. The geometric acceptance computed via this Monte Carlo approach was validated using the analytical approach described in Appendix~\ref{app:geometric_acceptance}.

\subsubsection{Tau Neutrino Propagation \& Tau Lepton Decay}

Tau neutrino propagation and $\tau$ lepton decay are treated using the NuTauSim code~\cite{Alvarez-Muniz2018}. This treatment includes tau neutrino and $\tau$ lepton propagation, regeneration, and the energy losses of the $\tau$ while traversing the Earth. Using NuTauSim, we construct lookup tables of the emerging $\tau$ lepton energies and exit probabilities over a full range of emergence angles and incident neutrino energies. We assume the mean cross sections from Ref.~\cite{Connolly2011} and the energy losses from Ref.~\cite{ALLM1997} in constructing the lookup tables.

Several results from Ref.~\cite{Alvarez-Muniz2018} are relevant to the high-elevation detector concept. First, the Earth acts as an energy spectrometer for $\nu_\tau$'s. Tau neutrinos with energies $> 10^{18}$~eV have exit probabilities that are several orders of magnitude higher at skimming emergence angles, $\theta_{\mathrm{emerge}}< 3^{\circ}$, than at steeper angles, while $\nu_\tau$'s with energies between $10^{15}$ and $10^{17}$~eV are equally probable to emerge from steep and skimming angles. Second, energy losses~\cite{ALLM1997,ASW2005}
and regeneration of neutrinos through neutral current interactions and $\tau$ decays result in a pileup of exiting $\tau$ lepton energies at $\mathcal{E}_{\tau}\sim10^{17}$~eV for emergence angles between $\theta_\mathrm{emerge} \simeq 0.3-3^{\circ}$. 
The two effects combined result in an expected enhanced flux of $\tau$ leptons with energies $10^{17}$~eV, so a detector tuned to this energy regime is preferred. 

\begin{figure}
    \centering
    \includegraphics[width=0.5\textwidth]{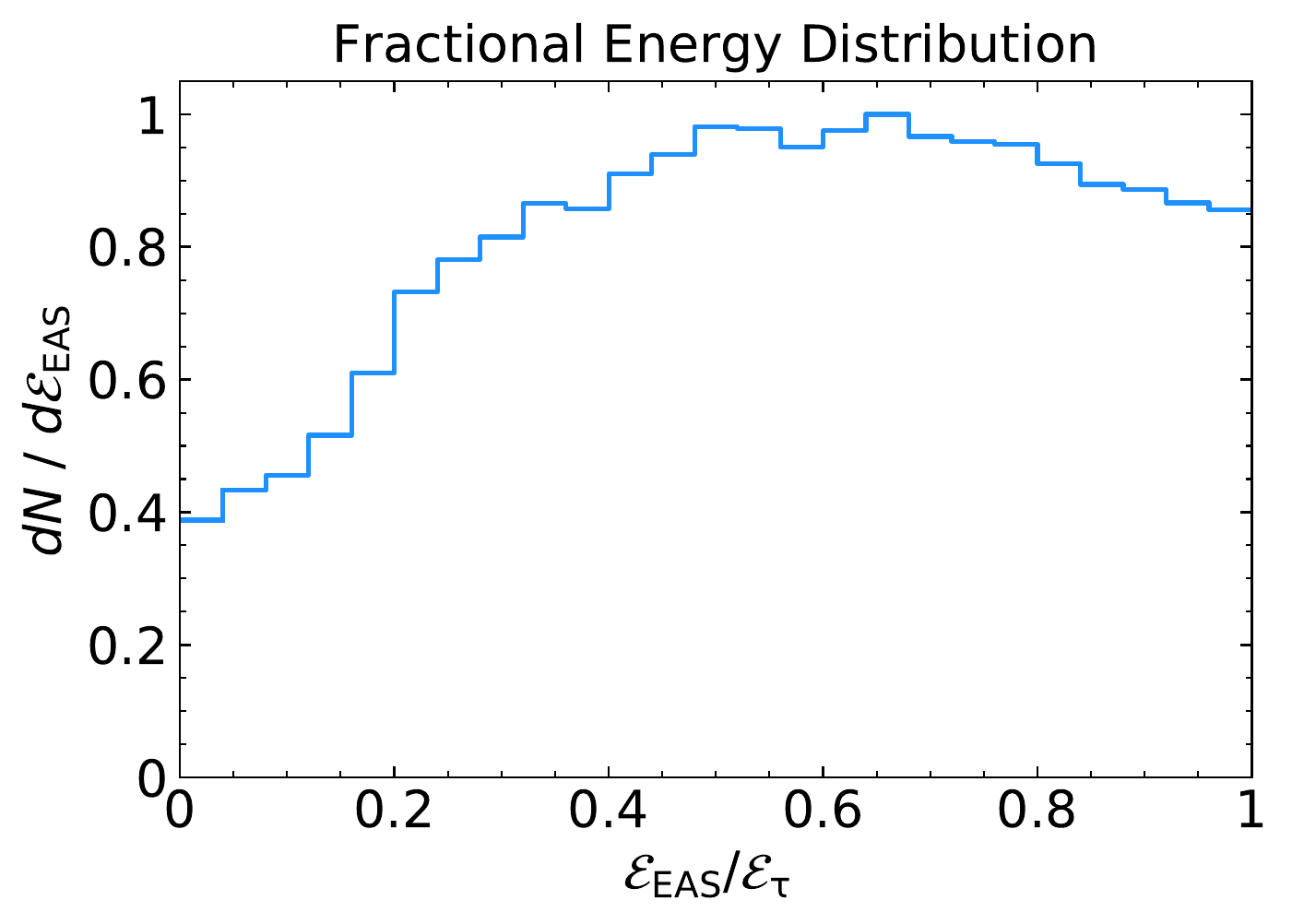}
    \caption{Distribution of the fraction of energy deposited into an extensive air shower from 10,000 decays of $\tau$ leptons of energy $10^{16}$~eV simulated with PYTHIA~\cite{Sjostrand2015}, normalized by the peak of the distribution. }
    \label{fig:decay_mode_energy_sampling}
\end{figure}

 Tau lepton decays result in an extensive air shower if the decay products include hadrons (branching ratio  $\Gamma >60$\%) or electrons  ($\Gamma\sim17.9$\%) or gamma rays. Only the showering components of the decay products  efficiently contribute coherent radio emission through multiple particles, while 
muons and neutrinos contribute zero or negligibly, effectively removing a fraction of the original $\tau$ lepton energy by carrying it far away.  In the Monte Carlo, the non-showering decays, like, \textit{e.g.}, $\tau \rightarrow \mu \nu_{\mu} \nu_{\tau}$ ($\Gamma\sim17.4\%)$, are treated by not allowing them to trigger the detector. For the showering decays, we estimate the shower energy, $\mathcal{E}_\mathrm{EAS}$, by sampling the negative helicity distribution of the $\tau$-lepton decays. We do this because the charged current interaction of a $\nu_\tau$ that produces the $\tau$ lepton is mediated by the W boson, which couples only to left-handed chirality and therefore negative helicity at ultra-high energies. Fig.~\ref{fig:decay_mode_energy_sampling} shows the distribution of the fraction of the $\tau$-lepton energy that is transferred to the air shower, computed using 10,000 $\tau$ lepton decays at ultra-relativistic energies with PYTHIA~\cite{Sjostrand2015}. The distribution shown has an average of 0.56 with a 68\% confidence interval of 60\%. 

Taking together the distributions of the emerging $\tau$-lepton energy from NuTauSim and the fraction of the energy transferred from the $\tau$ lepton to the shower, the combined median energy transferred from the neutrino to the shower at $10^{17}$~eV is 50\% with a  the 68\% confidence interval of 40\%. 

\subsubsection{Radio Emission}

The $\tau$-lepton induced extensive air showers showers generate radio emission that we model with ZHAireS, modified for treatment of upward-going showers. ZHAireS is a detailed Monte Carlo program capable of simulating extensive air showers induced by the products of a $\tau$ lepton decay as well as calculating its associated radio emission~\cite{Alvarez-Muniz2012}. We create look-up tables for each mountain elevation as follows. A line of detectors is placed along the mountain ridge corresponding to uniformly spaced view angles from the decay point ($\theta_{view}$ in Fig.~\ref{fig:geom_concept}). Each detector observes the time-domain electric field from the decay of a $\tau$ shower of $10^{17}$~eV via the dominant decay mode ($\tau \rightarrow \nu_{\tau} \pi^{0} \pi^{-}$), and we assume that the electric field scales linearly with shower energy~\cite{Glaser2016}. 

We simulated $\tau$-lepton induced air showers at several mountain elevations, emergence angles, decay altitudes, and view angles from the decay point and stored the peak electric fields in look-up tables. The parameters used in the look-up tables used in the main simulations are shown in Table~\ref{tab:zhaires_sims}. The atmospheric model used in the showers varies exponentially with altitude according to the U.S. Standard Atmosphere. The refractive index used in ZHAireS decreases exponentially with altitude, starting at 1.0003274 at sea level~\cite{Alvarez-Muniz2012}. After selecting the appropriate lookup table, the electric fields stored in the lookup tables are corrected by a linear scaling with shower energy and distance between the station and the decay point, $s_{\mathrm{RF}}$, for each event in the Monte Carlo.

Each ZHAireS simulation used to generate the electric field look-up tables has been simulated selecting the azimuthal angle of the neutrino arrival direction 
perpendicular to (a vertical plane containing) the magnetic field to maximize the radio signal obtained.
Such a design can be realized by careful consideration of the orientation of the array at a given site, but this assumption results in an optimistic estimate of the peak electric fields. 

\begin{figure}
\centerline{\includegraphics[width=\textwidth]{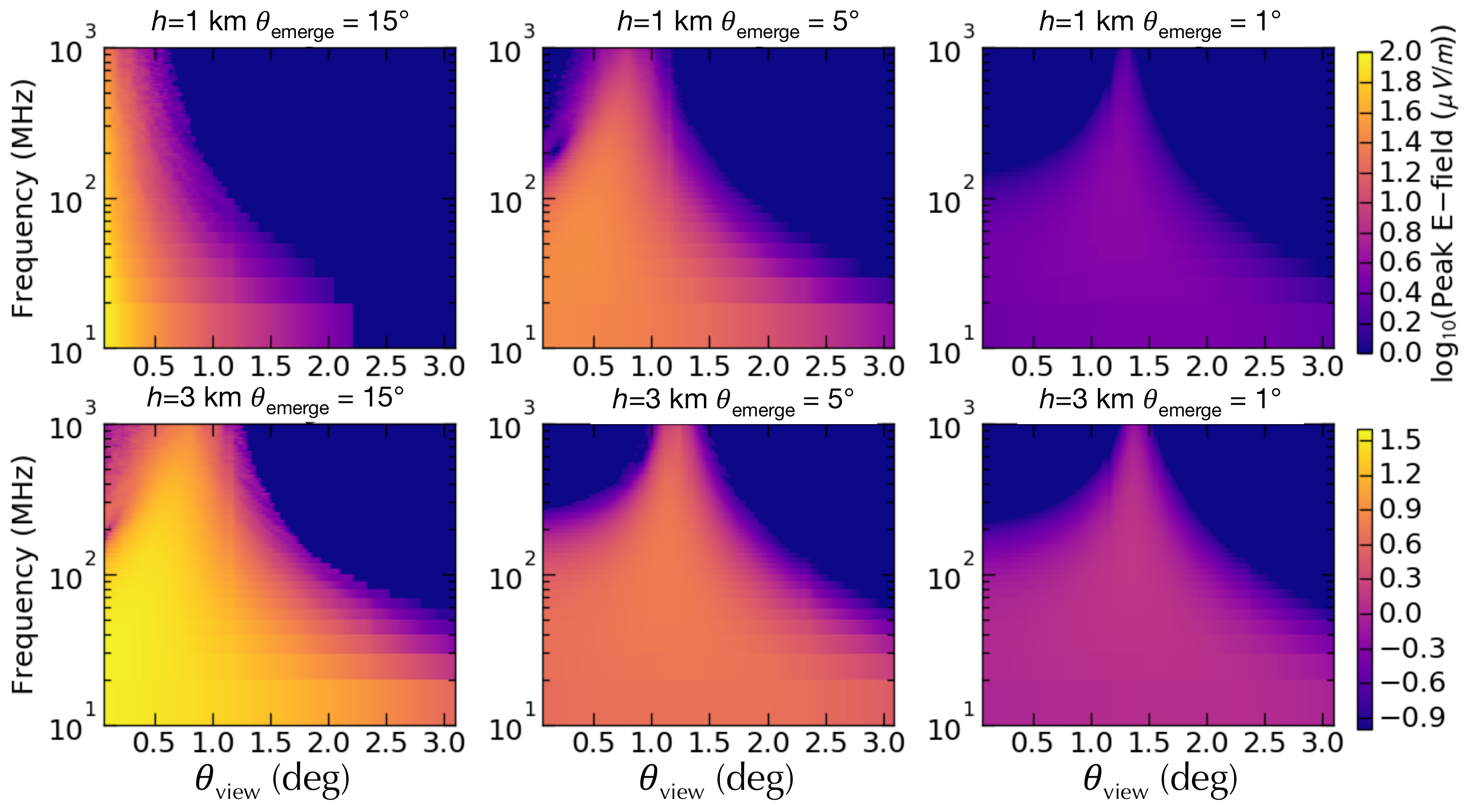}}
\caption{Peak electric fields from ZHAireS simulations of a $10^{17}$~eV $\tau$-lepton induced air shower decaying at ground level ($h=0$~km). Stations placed at elevations of $h=1$~km (top) and $h=3$~km (bottom) receive radio emission at emergence angles $\theta_\mathrm{emerge}$ of 15$^{\circ}$, 5$^{\circ}$, and 1$^{\circ}$. Each panel includes the peak electric field in 10-MHz wide sub-bands starting at the frequency indicated in the plot. A Cherenkov-like cone emerges when the distances between the shower and the detector are large, typically at high station elevations and/or small emergence angles, showing the transition from near-field to far-field.}
\label{fig:airshowerElectricField}
\end{figure}

The Fourier transform, $\tilde{E}(f)$, of the time-domain electric fields $E(t)$ registered at each antenna position is used to estimate the peak electric field for each 10~MHz subband with a center frequency $f_c$
\begin{equation}
    E_{\mathrm{peak}}(f_c) = 2 \int^{f_{c} + 5\,{\mathrm{MHz}}}_{f_c-5\,{\mathrm{MHz}}} |\tilde{E}(f)| df.
\end{equation} 
The peak electric fields as a function of $h$, $h_{decay}$, $\theta_\mathrm{emerge}$ and $\theta_{view}$ are then stored in look-up tables such as those shown in Fig.~\ref{fig:airshowerElectricField} for a decay altitude of 0~km above sea level (a.s.l.). 

When the shower is in the far field of the station as in the case of a 1 degree emergence angle as shown in the right panels of Fig.~\ref{fig:airshowerElectricField}, the signal is broad at low frequencies, and narrows to a sharp Cherenkov cone at higher frequencies. The signal strength scales linearly with distance to the shower for skimming showers of 1 degree emergence angles, but also for higher emergence angles if the station elevation is sufficiently high (see $\textit{e.g.},$ the bottom middle panel of Fig.~\ref{fig:airshowerElectricField} for station altitude $h=3$~km and $\theta_{\mathrm{emerge}}=5^{\circ}$). Steeper emergence angles or lower mountain elevations can result in geometries where the shower is not fully developed or the station lies within the near field of the shower. 

\begin{figure}
\centerline{\includegraphics[width=0.8\textwidth]{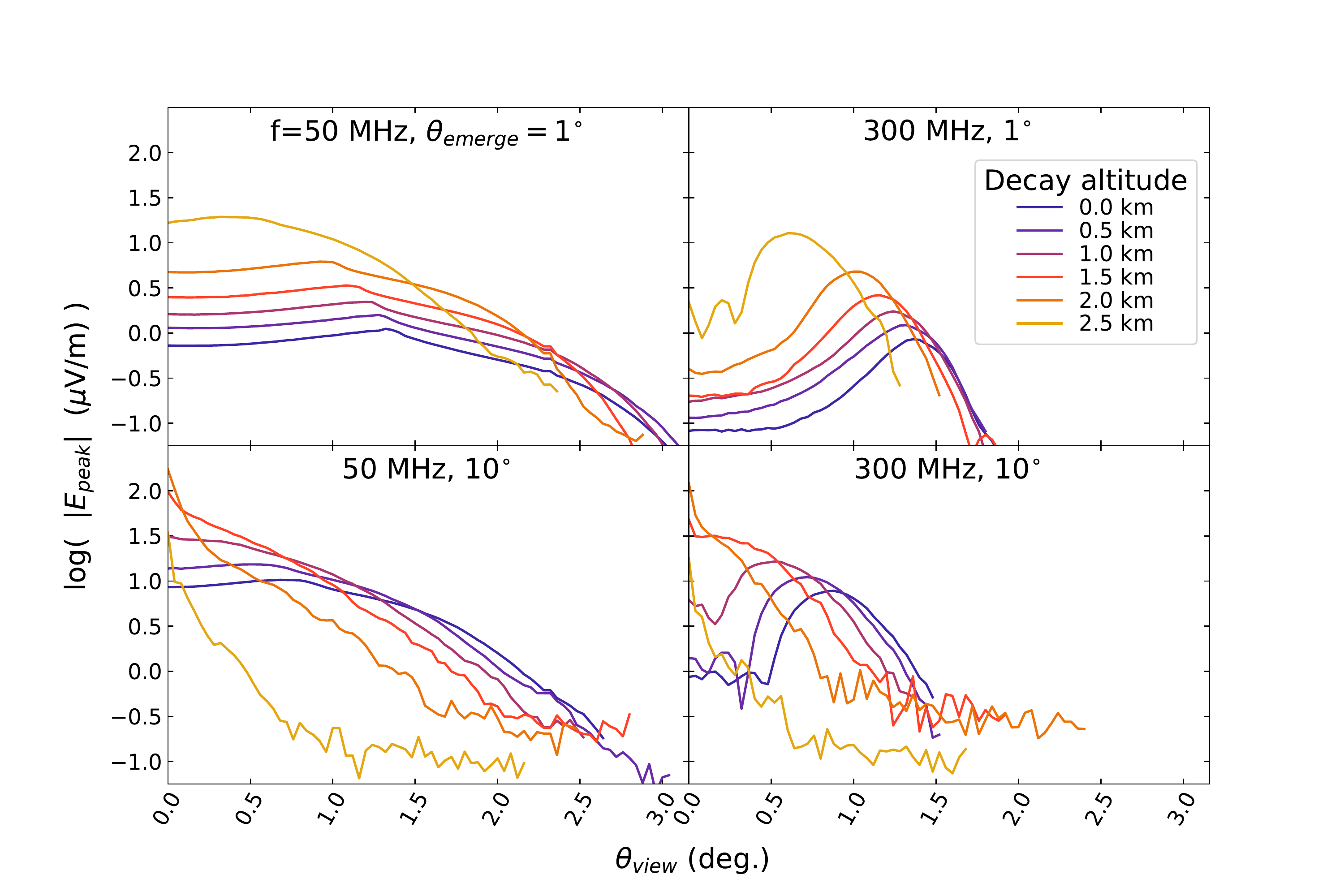}}
\caption{Peak electric field at 50 MHz (left) and 300 MHz (right) for a $10^{17}$~eV $\tau$-lepton induced shower for varying decay altitudes, at a fixed detector altitude of $h=3$~km and emergence angles of $\theta_\mathrm{emerge}=1^{\circ}$ (top) and $ 10^{\circ}$ (bottom). }
\label{fig:decayAltitude}
\end{figure}

Similarly, the shower moves into the near field of the station if the $\tau$ lepton decays at high altitude in the atmosphere. The radio beam patterns shown in Fig.~\ref{fig:decayAltitude} reflect the formation of a Cherenkov cone. A distinct Cherenkov ring is evident at higher frequencies due to coherence effects, while at lower frequencies, the beam pattern is broader. However, the peak electric field exponentially decreases with view angle if the $\tau$ decays close enough to the station so that the shower crosses the beamformer array as shown on the bottom panels. A linear scaling with distance is not sufficient to account for the electric field strength for decays high in the atmosphere or at large emergence angles, motivating the need for look-up tables as a function of decay altitude.

In this study we have assumed that the horizon is at sea level. However, our results apply to first order to any high-elevation detector site that has a wide view of the horizon and where the relative difference in elevation between the detector and the horizon is equivalent to $h$. If the horizon is higher than sea level, the acceptance is modified slightly because the showers evolve higher in the atmosphere, bringing the depth of shower maximum, $\chi_\mathrm{max}$, closer to the station~\cite{Romero-Wolf2019}. The acceptance of a given station will in general depend on the local topography~\cite{Decoene2019}.

\begin{table}
\begin{center}
\begin{tabular}{|c|c|}
    \hline
     Simulation Parameters &
     Values \\
          \hline
    
    Decay Mode & $\tau \rightarrow \nu_{\tau} \pi^{-} \pi^{0}$\\

     Primary $\tau$ Energy & $10^{17}$~eV\\
     Geomagnetic Field & 60$\,\mu$T, perpendicular to shower axis\\
     Zenith Angles at the decay point & 50, 55, 60, 65, 70, 75, 80, 85, 87, 89$^{\circ}$\\
     Decay Point Altitudes, $h_{\mathrm{decay}}$ & 0.5, 1.0, 2.5, 2.0, 2.5, 3.0, 3.5~km a.s.l.\\
     Detector Elevations, $h$ & 0.5, 1.0, 2.0, 3.0, 4.0~km a.s.l.\\
     View angles from the decay point, $\theta_{\mathrm{view}}$ & 0.0$^{\circ}$ to 3.2$^{\circ}$ in 0.04$^{\circ}$ increments \\
     \hline
\end{tabular}
\end{center}
\caption{Parameters of the suite of simulations run with ZHAireS to model the electric fields generated by upgoing $\tau$ lepton induced air showers. See Fig.~\ref{fig:geom_concept} and the text for definitions of the geometric parameters.}
\label{tab:zhaires_sims}
\end{table}

\subsubsection{Detector Model}
\label{Section:Detector}

The electric field incident at each antenna, $E_{\mathrm{inc}}$, generates a voltage at the feed in a ratio that depends on the antenna's effective height. The power received at the antenna, $A_{\mathrm{eff}} |E_{\mathrm{inc}}|^2 /2\eta$, scales with the square of the incident electric field, $E_{inc}$, and the antenna's effective area,
$A_{\mathrm{eff}}$. The impedance of freespace, $\eta$, is $377~\Omega$.  The effective area is given by $A_{\mathrm{eff}} = \frac{c^2}{4 \pi f^2} G(\theta, \phi)$, where  $c$ is the speed of light, $f$ is the frequency, and $G(\theta, \phi)$ is the gain of the antenna. The open circuit voltage measured at the feed of the antenna is then given by
\begin{equation}
    V_{\mathrm{o.c.}} =  \frac{2c}{f} \sqrt{\frac{G(\theta, \phi)}{4 \pi} \frac{R_A}{\eta} } |E_{\mathrm{inc}}|.
\end{equation}

The antenna and receiver form a voltage divider that splits the open circuit voltage at the feed based on the impedance of the load at the receiver, $Z_L$, and the antenna impedance, $Z_A = R_A + i R_X$. An impedance mismatch between the antenna and the receiver results in a loss due to power being reflected back to the antenna that depends on the reflection coefficient, $\Gamma = \frac{Z_L - Z_A^{*}}{Z_L + Z_A}$. Summing signals into beams formed from multiple antennas increases the receiver voltage by the number of antennas, $N$, in the center of the beam. The voltage measured at the receiver load from a beam of antennas is then

\begin{equation}
    V_{L} = \frac{2cN}{f} \frac{|Z_L|}{|Z_A + Z_L|} \sqrt{\frac{G(\theta, \phi)}{4 \pi} \frac{R_A}{\eta} (1-|\Gamma|^2)} |E_{\mathrm{inc}}|.
\end{equation}

To estimate the maximum performance, we assume a perfectly matched antenna where $Z_L = Z_A^{*}$ and a uniform gain $G(\theta,\phi)=G_0$. This allows us to study the range of zenith angles from which the $\nu_\tau$'s are likely to trigger in this geometry and orient the antennas and beams formed in the trigger accordingly. The peak voltage in a given 10-MHz subband with center frequency $f_c$ in the lookup tables is
\begin{equation}
V_{\mathrm{sub}}(f_c) \sim E_{\mathrm{peak}}(f_c) \frac{c}{f_c} N \sqrt{\frac{R_A}{\eta} \frac{G_0}{4\pi}}
\end{equation}
The full band voltage $V_{\mathrm{peak}}$ for a chosen detector design integrates the subband voltage $V_{\mathrm{sub}}$ across the passband. 

The noise at the receiver is generated by thermal noise due to the combination of the antenna temperature and the system noise temperature generated by  components of the electronics~\cite{Ellingson2005}. The antenna temperature depends on the fractional ratio of the sky to ground viewed by the antenna, $r$, and the galactic noise temperature, $T_{\mathrm{gal}}$, and the thermal noise temperature of the ground visible to the antenna, $T_{\mathrm{ground}}$. Because the average galactic noise temperature follows a power law in frequency, the noise temperature assumed in these simulations is dominated by the sky noise for the lower frequency band ($\lesssim 100$~MHz) and by the ground noise temperature at higher frequencies. We assume a typical ground noise $T_\mathrm{ground}=290$~K and a system noise $T_{\mathrm{sys}}=140$~K. For the sky temperature $T_\mathrm{gal}$, we use the Dulk paremetrization~\cite{Dulk_2001}. Phasing $N$ antennas sums the incoherent noise power 
such that the RMS voltage, $\sigma$ is
\begin{equation}
    \sigma = V_{\mathrm{RMS}} = \sqrt{ \int_{f_{\mathrm{lo}}}^{f_{\mathrm{high}}} N k R_L [r T_{\mathrm{gal}}(f)  +(1-r) T_{\mathrm{ground}} + T_{\mathrm{sys}}]  ~df},
\end{equation}
where $R_L$ is the real part of $Z_L$ and $k$ is the Boltzmann constant. On one antenna, the RMS voltage is 14$\,\mu$V in the 30-80~MHz band and 10$\,\mu$V in the 200-1200~MHz band, if we assume that $r$ is 0.5. In the following sections, thresholds in the trigger are referenced to these constant thermal noise levels in a given beam. 
 
\section{Reference Designs}\label{sec:refdesign}

By varying the number of antennas in the beamformer array, the frequency band, and the elevation, the results of the Monte Carlo acceptance estimates lead us to two reference designs. Each detector consists of 100 stations  with a beamformer array that phases together 10 antennas at a 3~km prominence with a 120 degree view in azimuth of the nearby horizon in either a lower frequency band (30-80~MHz) or a higher frequency band (200-1200~MHz) using commonly available antennas in each frequency band. We assume a single antenna gain of $G_0=1.8$~dBi for the lower frequency band, corresponding to an electrically short, dipole antenna. We assume a single antenna gain of $G_0=10$~dBi for the higher frequency band, corresponding to horn or log periodic dipole antennas.  

\begin{figure}
\includegraphics[width=\textwidth]{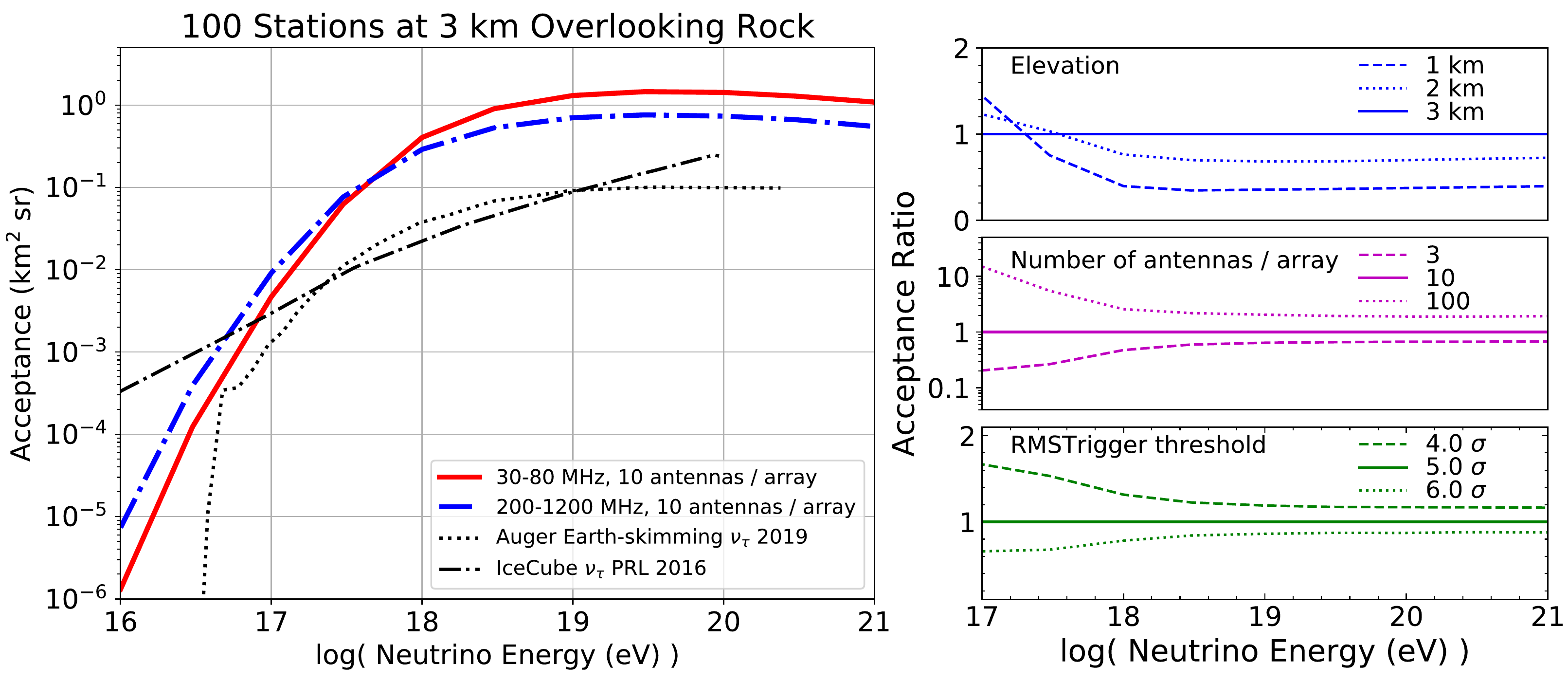}
\caption{(Left) Acceptance for 100 stations of the two reference designs: (red, solid) 30-80 MHz, 10 antennas in the beamformer array, $G_i$=1.8~dBi, $G_\mathrm{eff}$=11.3~dBi, 3~km elevation, 5$\sigma$ threshold; (blue, dot-dashed) 200-1200 MHz, 10 antennas, $G_i$=10~dBi, $G_\mathrm{eff}$=20~dBi, 3~km elevation, 5$\sigma$ threshold.
 The acceptances are compared to those from Earth-skimming $\nu_\tau$'s in Auger~\cite{Auger2019} and $\nu_\tau$'s in IceCube~\cite{IceCube2016}. (Right) The ratio of the acceptance relative to the reference design of a 30-80~MHz design for three different elevations (top), for arrays having different numbers of antennas (middle), and for three different voltage trigger thresholds in the beams: 4$\sigma$, 5$\sigma$, and 6$\sigma$ (bottom). The trigger levels are referenced to thermal noise in the 30-80 MHz band of 14~$\mu$V~$\sqrt{N}$. 
}
\label{fig:refdesignacceptance}
\end{figure}

The two reference designs achieve the acceptances shown in Fig.~\ref{fig:refdesignacceptance} (left). Note that these refined estimates of the acceptance that include detailed modeling of the electric fields and trigger are within an order of magnitude of the first-order, maximal estimates from Section~\ref{sec:firstorder}. Because high-elevation stations are mainly sensitive to $\nu_\tau$'s, we compare to the acceptance of Auger to Earth-skimming $\nu_\tau$'s~\cite{Auger2019} and and to that of IceCube to $\nu_\tau$ cascade events~\cite{IceCube2016}. Both reference designs improve on the acceptance of existing experiments for energies greater than $10^{17}$~eV and can reach more than a factor of 10 improvement at the highest energies. 

The lower frequency band achieves a higher acceptance at energies greater than $3\times10^{17}$~eV, while the higher frequency band has a lower energy threshold. The differences can be attributed to the differences in the radio beam patterns shown in Figs.~\ref{fig:airshowerElectricField} and \ref{fig:decayAltitude}. The lower frequency radio beams are broader, since the region over which the radio emission is coherent is broader at longer wavelengths. However, the peak electric field measured in the higher frequency band is stronger at the Cherenkov angle due to the increase in bandwidth and lower galactic noise contribution.  Given that the two reference designs differ in acceptance by at most a factor of 2, the choice of the frequency band may be better determined by the radio-frequency interference local to a given site or the requirement on cosmic ray rejection.

\begin{figure}
\includegraphics[width=\textwidth]{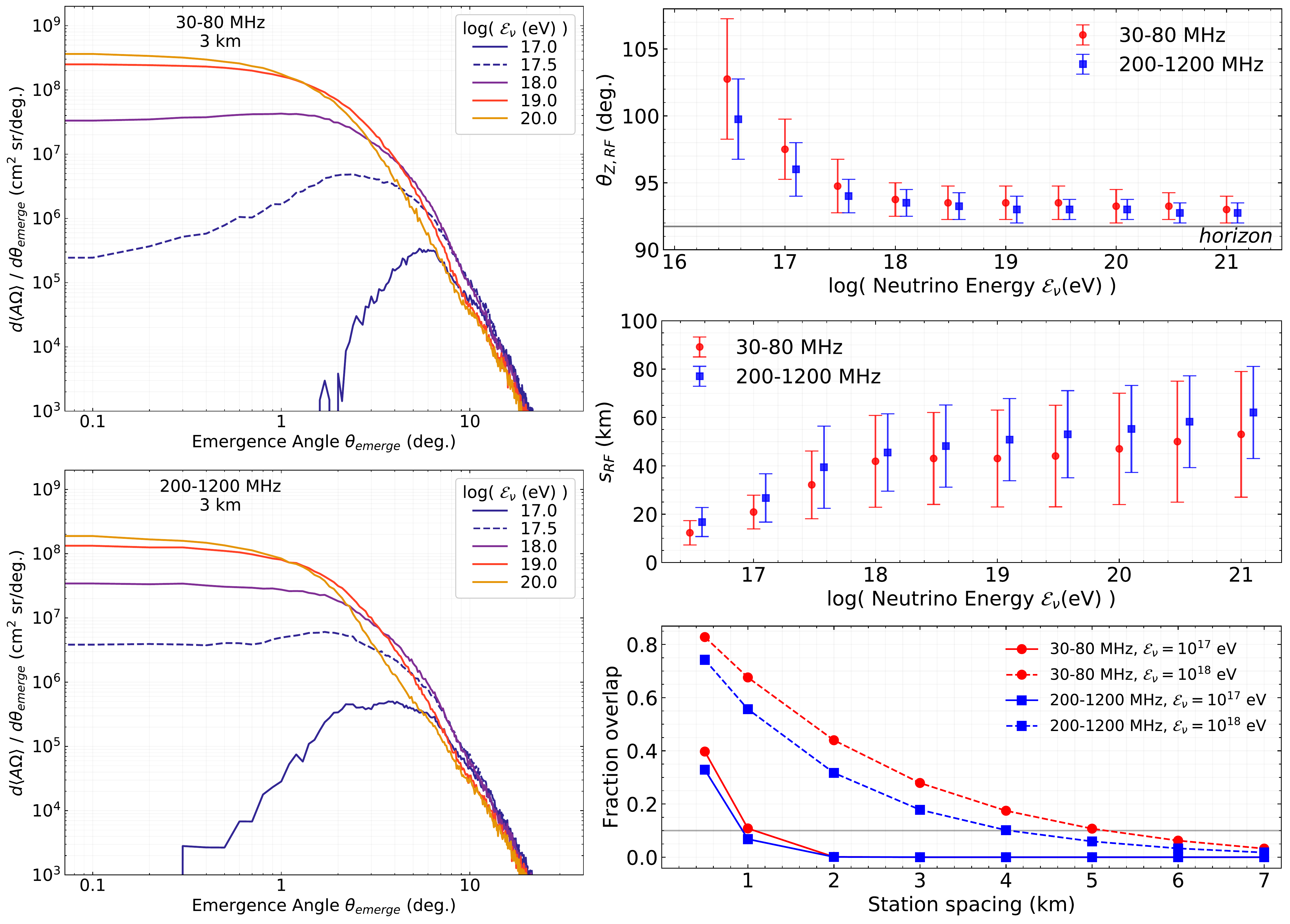}

\caption{ 
Differential acceptance and triggered event distributions for the two reference designs: (top left and red circles on right) 30-80 MHz, 10 antennas in the beamformer array, $G_i$=1.8~dBi, $G_\mathrm{eff}$=11.3~dBi, 3~km elevation, 5$\sigma$ threshold; (bottom left and blue circles on right) 200-1200 MHz, 10 antennas, $G_i$=10~dBi, $G_\mathrm{eff}$=20~dBi, 3~km elevation, 5$\sigma$ threshold. (Left) Single station differential acceptance dependence on emergence angle at the exit point $\theta_\mathrm{emerge}$. (Top right) The median and 68\% confidence intervals of the distribution of $\theta_{Z,RF}$ and (middle right) the distance between the decay point and the detector, $s_\mathrm{RF}$, for triggered events. The blue points are shifted by $\log{\mathcal{E}_{\nu}}$ = +0.1 for legibility. (Bottom right) The fraction of events triggering two stations separated by a given station spacing at two different energies ($10^{17}$~eV and $10^{18}$~eV). See Fig.~\ref{fig:geom_concept} for geometric definitions.
}
\label{fig:diffacceptance}
\end{figure}

The mountaintop detector concept is most sensitive to events emerging near the horizon.
The left panel of Fig.~\ref{fig:diffacceptance} shows the differential acceptance for the two reference designs for varying emergence angles.  At the highest energies $>3\times10^{17}$~eV, the differential acceptance peaks at emergence angles below a few degrees and is nearly flat closer to the horizontal. At lower energies, the showers need to be closer to trigger the detector, typically within 
$\sim50$~km,
(see the middle right panel of Fig.~\ref{fig:diffacceptance}), 
such that $\nu_\tau$ events emerge at angles larger than a few degrees. 

The zenith angles of the radio emission received at the station $\theta_{Z,RF}$ (see Fig.~\ref{fig:geom_concept}) from triggered events are shown on the top
right panel of Fig.~\ref{fig:diffacceptance}. The median zenith angles of the lower frequency design are more than 2 degrees from the horizon at all energies, while the median zenith angles for the higher frequency design are closer to the horizon. Events in the lower frequency design trigger over a broader range of angles, broadening the width of the distribution. These results suggest that a trigger design based on the lower frequency design is preferred for cosmic ray background rejection, because the showers are expected to come upwards with larger nadir angles. Outrigger antennas could use the higher frequency band for better pointing resolution and improved cosmic ray rejection. Additionally, based on the separation between the triggered events and the horizon, a pointing resolution of better than 1 degree is required for the successful background rejection of down-going cosmic rays.

The mountaintop detector may include multiple stations along the same ridge, multiple stations placed at different sites around the world, or both. To ensure that each station views an independent portion of the nearby horizon, the stations need to be separated by enough distance that events rarely trigger multiple detectors. The bottom right panel of Fig.~\ref{fig:diffacceptance} shows that the required station spacing depends on the chosen frequency band and energy. Stations would need to be spaced by at least 5~km to ensure that less than 10\% of the events trigger two adjacent detectors for energies less than $10^{18}$~eV.

\section{Design Variations}\label{sec:design_vars}

The performance of a high-elevation $\nu_\tau$ detector is impacted by several design choices, often driven by characteristics of a given site. We consider three design variations in the right panel of Fig.~\ref{fig:refdesignacceptance} for the 30-80 MHz design in terms of the elevation, the gain of the array (i.e. number of beamformed antennas), and the trigger threshold, which depends on the RFI environment of the array. 

Stations at higher elevation view a larger area and are far away enough from the showers that the $\tau$ lepton can decay and the air shower can fully develop. This results in a higher acceptance at energies greater than $3\times10^{17}$~eV as the stations are placed at higher elevation, with the improvement in acceptance being larger going from 1~km to 2~km than from 2~km to 3~km as shown on the top right panel of Fig.~\ref{fig:refdesignacceptance}. At lower neutrino energies, each  station is more likely to trigger on events with a shorter distance between the shower and the station, resulting in a 50\% higher acceptance of the detector placed at 1~km over one placed at 3~km. These results suggest that mountain ridges at least 2~km above the horizon should be considered for a detector of this design, representing a good compromise between maximizing the acceptance at the highest energies, minimizing the number of required stations, and maintaining a low energy threshold of $10^{17}$~eV.

As can be inferred by the middle right panel of Fig.~\ref{fig:refdesignacceptance}, phasing more antennas can impact the acceptance at high energies and the energy threshold of the detector, because it increases the effective phased antenna gain $G_{\mathrm eff} = G_{0} + 10 \log(N)$ where $G_{0}$ is the gain of a single antenna in the array. The acceptance at $10^{17}$~eV increases by an order of magnitude with each factor of ten increase in the number of antennas used in the array. However, at energies higher than $10^{18}$~eV, the improved gain when going from 10 antennas (11.8 dBi) to 100 antennas (20.0~dBi) results in only a factor of 2 increase in the acceptance. This suggests that the optimal number of antennas to include in the beamformer array is $\sim10$. A ten-antenna array can achieve a lower energy threshold and higher acceptance overall compared with a three-antenna array, but also improved pointing resolution and hence background rejection. Beyond 10 antennas, it would be better to build more stations viewing independent regions of the horizon than to add more antennas to each station, because the acceptance scales linearly with the number of stations.

The trigger threshold of an impulsive radio receiver is implemented as a noise-riding threshold that adjusts to meet a predefined global trigger rate, typically between 1-100 Hz. Fluctuations in the local RFI environment due to anthropogenic backgrounds can cause the thresholds to vary over time and in different beams. The bottom right panel of Fig.~\ref{fig:refdesignacceptance} shows that small variations in the assumed trigger threshold can impact the acceptance. Lower average trigger thresholds will result in improved acceptance at all energies and a lower energy threshold, while higher trigger thresholds  have the opposite effect. We conservatively assume a voltage trigger threshold of 5$\sigma$ in the beams based on the successful implementation of the phased array technique in an ARA station~\cite{oberla_phasedarray_ara}. 

\begin{figure}[th]
\centerline{\includegraphics[width=\textwidth]{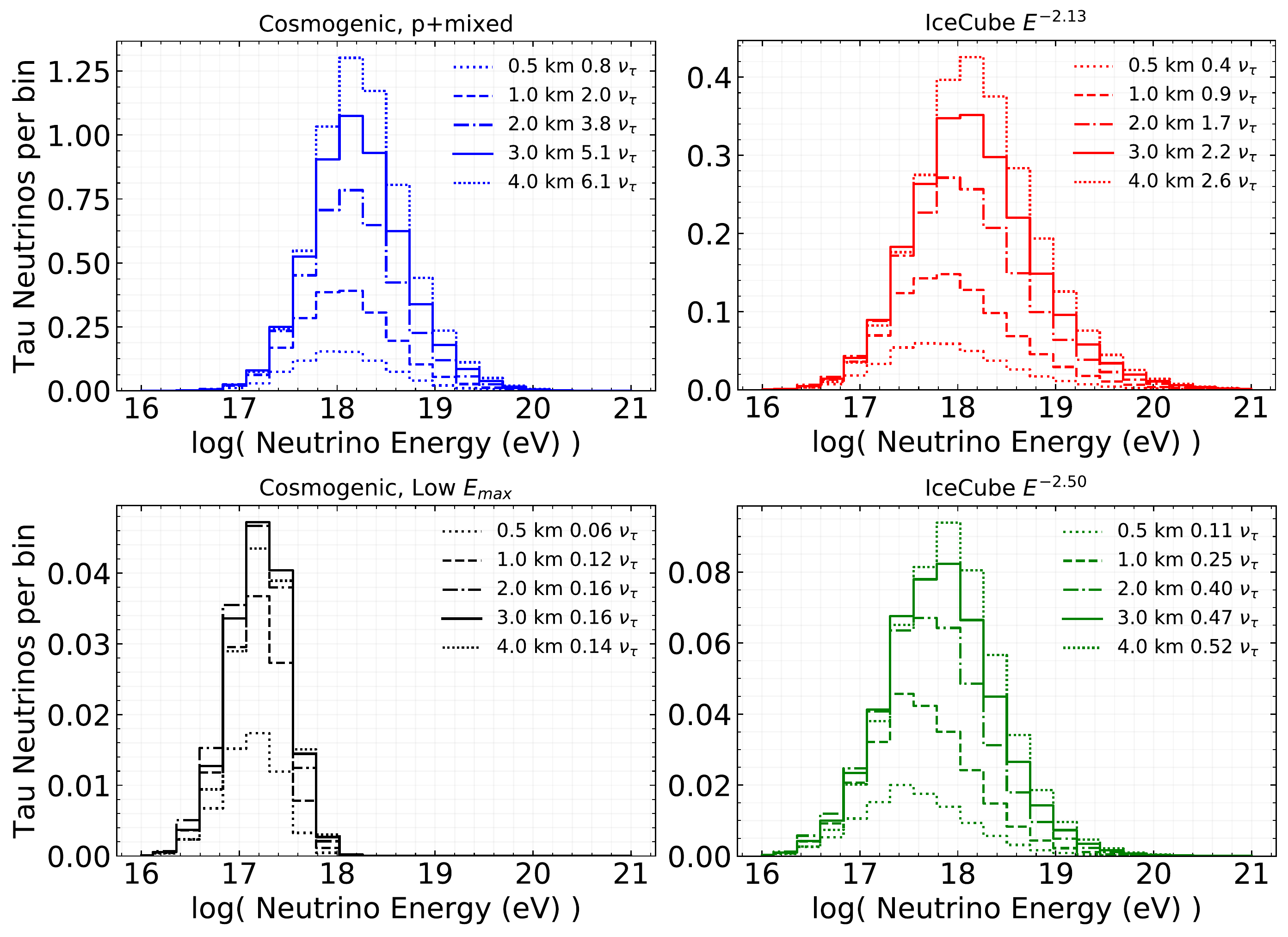}}

\caption{ Expected number of neutrino events per quarter-decade energy bin observed in 5 years with 100 stations, using the 30-80 MHz reference design placed at four elevations. Four different models bracket the possible cosmogenic and astrophysical fluxes.  The integrated number of observed neutrinos for each assumed flux model and elevation is shown in the legend. 
Typical cosmogenic models are shown on the left panels such as one assuming a range of cosmic ray compositions (pure proton or a mixed compositions), transition models, and source distribution models (mean of the grey band in Fig. 9 in Ref~\cite{Kotera2010}, top left, blue) and another assuming consistency with Auger cosmic ray data above the GZK sphere resulting in a low source energy cutoff~\cite{Romero-Wolf2018} (bottom left, black). Shown on the right are neutrino energy distributions expected from the extrapolations of two different spectra measured by IceCube, the muon track analysis~\cite{IceCube2016a} (top right, red) and the all flavor analysis~\cite{IceCube2015} (bottom right, green). }

\label{fig:design_variations_energy}
\end{figure}
                
\begin{figure}[t]
\centerline{\includegraphics[width=\textwidth]{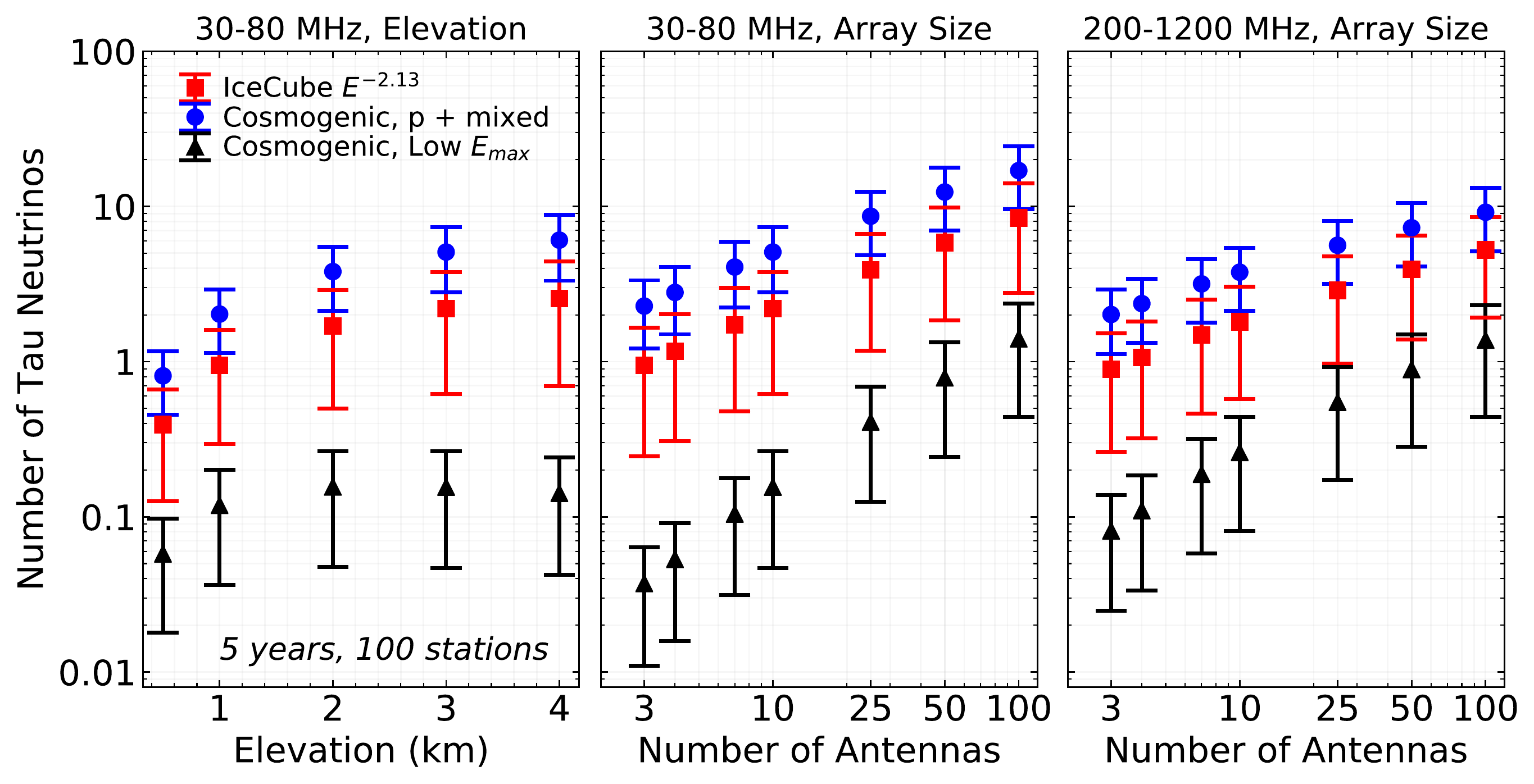}}
\caption{
Number of detected $\nu_{\tau}$'s for different elevations (left) and different number of antennas in the beamformer array (middle and right). The middle and right panels compare the performance of the lower and higher frequency reference designs. In the left panel, we assume 10~antennas in the beamformer array using the lower frequency design and in the middle and right panels, we assume a detector elevation of 3~km. 
 Colors and models assumed are the same as in Fig.~\ref{fig:design_variations_energy}. For the cosmogenic flux models that assume a proton and mixed cosmic ray composition (blue circles), the upper and lower error bars correspond to the upper and lower fluxes in the grey band of Fig.~9 of Ref.~\cite{Kotera2010}. For the cosmogenic low $E_\mathrm{max}$ fluxes (black triangles), the upper and lower bound correspond to the 68\% confidence interval in the posterior distribution of flux curves given the constraints on cosmogenic model parameters~\cite{Romero-Wolf2018}. The IceCube flux (red squares) is extrapolated to higher energies within the error band given by the power law fit~\cite{IceCube2016a}. }
\label{fig:design_variations}
\end{figure}

A mountaintop radio beamformer can constrain the flux of $\nu_\tau$'s at energies above $10^{17}$~eV with different effectiveness depending on how the spectrum behaves. Figs.~\ref{fig:design_variations_energy} and \ref{fig:design_variations} illustrate how the choice of detector design depends on the assumed neutrino flux model. We assume four different models that bracket different assumptions about the isotropic neutrino flux. Measurements of the isotropic neutrino flux are consistent with power laws up to a few PeV, but the spectral index varies depending on the event selection and analysis techniques. Through-going muon events favor a harder spectrum $dN_{\nu}/d\mathcal{E}_{\nu} \propto \mathcal{E_\nu}^{-\gamma}$ with a spectral index $\gamma=2.13\pm0.13$~\cite{IceCube2016a}, while analysis using the all-flavor and high-energy events contained within the detector favor softer spectra of $\gamma\sim2.5$~\cite{IceCube2015}. 

Cosmogenic models fall into two general categories based on the neutrinos~\cite{Berezinsky_Zatsepin_1969} expected from the Greisen-Zatsepin-Kuzmin (GZK) interaction of ultra-high energy cosmic rays~\cite{Greisen1966a, Zatsepin-Kuzmin-1966}. Some assume all possible cosmic ray compositions (proton dominated, mixed, or iron dominated) and many possible cosmic ray source distributions (e.g., that which follows the star formation rate). We use the mixed composition and proton dominated scenarios from Ref.~\cite{Kotera2010} as a representative example. Other cosmogenic models assume a low  cut-off energies ($\sim10^{17}$~eV), motivated by fits to the cosmic ray spectrum and composition measured by Auger. As a test case for those models we assume Ref.~\cite{Romero-Wolf2018}. 

The neutrino energy distribution of the triggered events for four detector configurations are compared in Fig.~\ref{fig:design_variations_energy} for four different assumptions of the diffuse neutrino flux. The models that extend to higher energies predict a factor of 6 or more neutrinos by moving the detector from a 0.5~km elevation to a 4~km elevation. If instead the $\nu_\tau$ flux is more suppressed at energies above $10^{17}$~eV as shown in the bottom row in Fig.~\ref{fig:design_variations_energy}, there is little benefit to building stations at elevations higher than 1~km. The models shown on the bottom row with either softer spectra or energy cutoffs result in an energy distribution of $\nu_\tau$'s that peaks towards the lower end of the sensitivity of the high-elevation detector. 

Fig.~\ref{fig:design_variations} compares the number of detected $\nu_\tau$'s in the Monte Carlo as the elevation, phased antenna gain, and frequency band change relative to the reference design. The number of $\nu_\tau$'s triggered in the Monte Carlo grows with elevation according to the expectation from the first-order estimates of the acceptance (proportional to $h^{3/2}$) for models that extend to higher energies.  To minimize the observation time required to discover a flux of $\nu_\tau$'s or constrain the models that extend to higher energies, these results suggest that 100 stations should be placed at a minimum elevation of 2~km using the 30-80~MHz reference design. The same detector placed at 3~km could observe 30\% more neutrinos, requiring that much less observation time. If no neutrinos are observed, then subsequent stations could be placed at a lower elevation (1-2 km), given that the models that are more suppressed at high energies (have softer spectra) show little variation with elevation in the number of observed neutrinos above 1~km.

For the models with harder spectra, the number of neutrinos grows roughly linearly with phased antenna gain (or the log of the number of antennas), but each factor of 2 increase in antenna gain requires a factor of ten increase in the number of antennas per beamformer array. At a minimum, three antennas are required to reconstruct the arrival direction. Including more antennas in the station improves direction reconstruction, but given that the number of neutrinos grows linearly with the number of stations, we conclude that beyond 10 antennas, it would be better to build more stations than make a larger beamformer array at each station. The lower frequency design registers more neutrinos with an array of 10 antennas for the models that extend to higher energies. 

In contrast to the results for the harder spectrum models, the softer spectra require a factor of ten more antennas to detect an appreciable number of neutrinos in 5 years. The number of detected neutrinos is unchanged as elevation increases, because the increase in visible area is balanced by the decrease in the fraction of nearby showers. The higher frequency reference design registers more neutrinos per antenna from the lower energy models, because the electric field strength at the Cherenkov cone is stronger. 

\begin{figure}[t]
\centerline{\includegraphics[width=\textwidth]{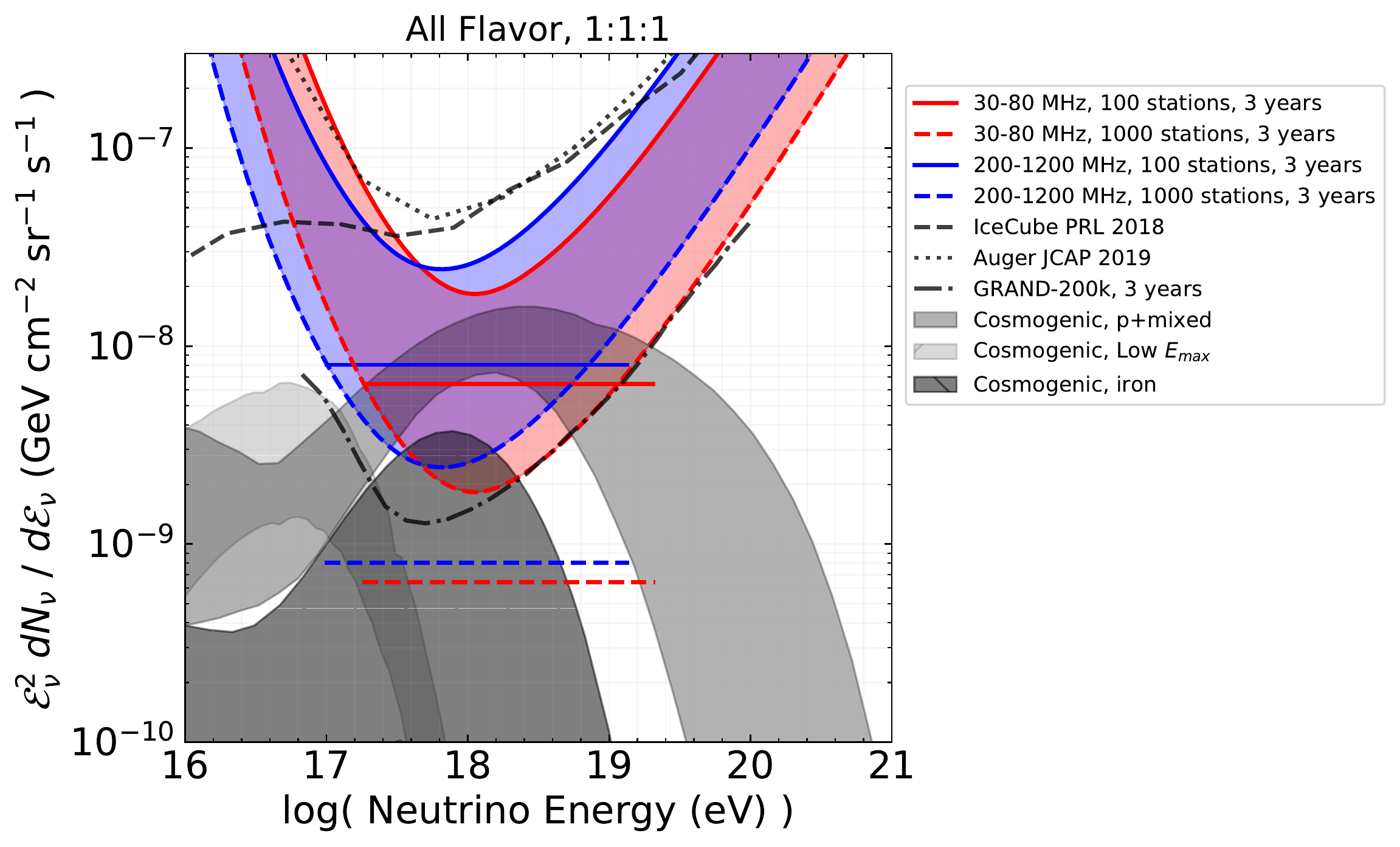}}
\caption{Sensitivity of a high-elevation mountaintop beamforming radio array to an $dN_{\nu}/d\mathcal{E}_{\nu} \propto \mathcal{E_\nu}^{-2}$ flux of neutrinos, assuming a 1:1:1 flavor ratio, half-decade energy bins, and a three year exposure. The expected 90\% confidence upper limit on the differential flux for 100 stations and 1000 stations of both the low- and high-frequency reference designs are compared with models of the cosmogenic flux~\cite{Kotera2010, Romero-Wolf2018, Kampert2012} and upper limits from Auger~\cite{Auger2019}, IceCube~\cite{IceCubeUHE2018}, and the proposed GRAND-200k experiment~\cite{GRAND_whitepaper}. For the cosmogenic flux models that assume a proton and mixed cosmic ray composition, the upper and lower error bars correspond to the upper and lower fluxes in the grey band of Fig.~9 of Ref.~\cite{Kotera2010}. For the cosmogenic low $E_\mathrm{max}$ fluxes, the upper and lower bound correspond to the 68\% confidence interval in the posterior distribution of flux curves given the constraints on cosmogenic model parameters~\cite{Romero-Wolf2018}. The cosmogenic fluxes that presume an iron composition are bounded by the model that assumes a pure iron and FRII evolution (Fanaroff-Riley type II AGN)~\cite{Kampert2012} and the minimum flux assuming an iron or iron-rich cosmic ray composition from Ref.~\cite{Kotera2010}. The expected integrated flux upper limits (horizontal lines) are shown in the energy region that includes 90\% of the expected events from an $\mathcal{E_\nu}^{-2}$ flux.}
\label{fig:sensitivity}
\end{figure}

The sensitivity of the reference designs to an all-flavor $dN_{\nu}/d\mathcal{E}_{\nu} \propto \mathcal{E_\nu}^{-2}$ flux is presented in Fig.~\ref{fig:sensitivity}, both as a differential upper limit in half-decade bins ($\Delta \log{\mathcal{E}_\nu} = 0.5$) and as an integrated flux upper limit. Assuming an exposure time, $T$, number of stations $N_{s}$, and a Feldman-Cousins factor~\cite{FeldmanCousins1997} for observing zero candidate and background events ($N_{90} = 2.44$), the projected all-flavor 90\% confidence limit, $k_{90}$, is given by:
\begin{equation}
    \mathcal{E}_\nu \, k_{90} = \frac{3\,N_{90}}{N_{s} \, \langle A \,\Omega \rangle \, T \, \ln{10} \, \Delta\log_{10}{\mathcal{E}_\nu}}.
\end{equation}
The integrated flux upper limit summed over $n$ logarithmic energy bins is $(\Sigma_{i=0}^{n} {\mathcal{E}_\nu^{2} k_{90}} )^{-1}$ and is shown in Fig.~\ref{fig:sensitivity} over the energy range where 90\% of the events would be expected from an $\mathcal{E}_{\nu}^{-2}$ flux for comparison to other experimental designs.

Assuming a three year exposure, the mountaintop detector can improve on the nearly 10-year limits from existing experiments~\cite{Auger2019,IceCubeUHE2018} by a factor of three above $10^{17}$~eV, consistent with the factor of 10 improvement in the acceptance and reduced observation time. 

It is also relevant to compare to the proposed GRAND experiment which plans to use 200,000 antennas in both the valley and on mountains to map out the radio emission from upgoing $\tau$-lepton induced showers~\cite{GRAND_whitepaper}. By using beamforming at high elevations, it is possible to match the expected GRAND sensitivity with 1000 stations, with the caveat that the stations need to be separated by enough distance to be sensitive to independent sets of upgoing $\tau$ showers. Since each beamformer array consists of 10 dual-polarized antennas, the total number of antennas (channels) required for the trigger would be 10,000 (20,000).

\section{Discussion}\label{sec:discussion}

A mountaintop radio beamformer discussed here can be an efficient detector for upgoing $\nu_\tau$'s. The design is predicated on a few assumptions including reasonably low background levels, a clear view of the horizon over a 120$^{\circ}$ field-of-view, and efficient radio-only triggering on inclined air showers. Once the tau-neutrino induced showers are detected, the energy and arrival direction of the incident neutrino will need to be reconstructed. We discuss these topics here.

The mountaintop radio detector design assumes that a site is radio quiet enough to trigger efficiently on signals $5\sigma$ above the thermal noise background over a 120 degree field of view in azimuth. This technological capability has yet to be demonstrated, but analyses from the TREND~\cite{Ardouin2011} and OVRO-LWA~\cite{Monroe2019} experiments suggest that radio-only triggering may be possible by rejecting triggers arriving from certain directions at the trigger level. A prototype beamformer array of four 30-80 MHz antennas has been deployed at the White Mountain Research Station in Bishop CA. using a phased array trigger~\cite{Hughes_ICRC_2019} which will be the subject of a subsequent study.

More advanced triggering or reconstruction schemes may also be possible. For instance, the trigger could be based on the lower frequency designs, but each station could also include higher frequency channels in the pointing array to further constrain the direction of the air shower. 
Given that the showers detected via their radio emission typically  originate from 10 to 80~km from the detector, a trigger that detects the upward-going air-shower muons using scintillators may have similar acceptance, but be less susceptible to man-made radio backgrounds. 

After designing a trigger that efficiently rejects any man-made backgrounds or choosing a site that minimizes them, the predominant physics backgrounds observed in a mountaintop detector will come from conventional down-going cosmic rays. The $10^{17}$~eV energy threshold of a mountaintop detector sets the expected rate of downgoing cosmic rays.  Assuming a station layout with 10 low frequency antennas arranged in a 70~m square grid on a $20^{\circ}$ slope at 3~km altitude, we sample the cosmic rays triggered in a station using the detector model described in Section~\ref{sec:detector} and parameterizations of the electric fields expected at the LOFAR site~\cite{Alvarez-Muniz:2014wna}. The distribution of zenith angles peaks at 30$^{\circ}$, and a small fraction ($2\times10^{-6}$) originates from within two degrees above the horizon. We estimate that the near horizon cosmic ray rate is 0.02 events per station in 3 years.  Radio signals from cosmic ray air showers that reflect off the ground undergo a phase inversion (in addition to attenuation due to the non-specular reflection), meaning that reflected cosmic ray events near the horizon can be distinguished from upgoing tau neutrinos via their polarity and radio spectrum. For a similar experiment, the probability that an air shower's polarity is mis-reconstructed has been shown to be $10^{-4}$ to $10^{-2}$~\cite{Gorham2016, Gorham2018, ANITA_nearhorizon}. Therefore, given a 1~degree radio pointing resolution (see top right panel of Fig.~\ref{fig:diffacceptance}), we expect the cosmic ray backgrounds to be negligible, motivating the background rate assumed in Fig.~\ref{fig:sensitivity}.

The first events detected by a mountaintop station will be conventional down-going ultra-high energy cosmic rays arriving from shallow zenith angles, far from the horizon.  These events can serve as a template predicting the pulse shape due to air showers. Because the mountaintop detector is stationary and continuously monitors the horizon, these backgrounds can be fully characterized within months of integration time. Candidate $\nu_\tau$ events can be scrutinized using detailed follow-up studies of the exact geometry using calibration signals on a drone~\cite{Nam_ICRC_2019}, following the work done with the HiCal calibration instrument accompanying ANITA~\cite{HiCal12017}, but with a more thorough range of reflection geometries, because the instrument is stationary. Anthropogenic backgrounds are expected to be repetitive signals, which most often occur with different spectral characteristics than air showers.

The neutrino arrival direction resolution depends on both the arrangement of the antennas in the reconstruction array and the frequency band chosen for the detector. At lower frequencies, the radio signal from the air shower is coherent over a broader range of angles ($\sim2^{\circ}$,  see Figs.~\ref{fig:airshowerElectricField} and \ref{fig:decayAltitude}), meaning that $\theta_{view}$ can only be reconstructed with comparable uncertainty. At higher frequencies, $\theta_{view}$ can be reconstructed using the radio spectrum with sub-degree resolution~\cite{Schoorlemmer2016}. There is a degeneracy in determining which side of the Cherenkov cone the radio signal arrives from. This suggests that the resolution on the neutrino arrival direction could be as broad as the Cherenkov angle summed with the uncertainty on the radio pointing, $\delta \theta$, or 3$^{\circ}$, depending on the layout of the reconstruction array.

As with all instruments that search for upgoing tau neutrinos, the reconstructed energy of the shower is only a lower bound on the true neutrino energy. Spectral unfolding techniques will be necessary to reconstruct the underlying spectrum. The energy resolution is ultimately limited by the unknown transfer of energy from the $\nu_\tau$ to the emerging $\tau$, which includes the $\tau$ energy losses during propagation (see Ref.~\cite{Alvarez-Muniz2018}), and the transfer of energy from the $\tau$ to the shower (see Fig.~\ref{fig:decay_mode_energy_sampling}). Radio techniques used in cosmic ray experiments have demonstrated that the shower energy can be reconstructed to 17-25\%, depending on the detector design~\cite{AERA_Energy_Reco:2015vta,Aab:2016eeq, Schoorlemmer2016}. To reconstruct the shower energy $\mathcal{E}_{rec}$, one must first reconstruct the distance to the decay $s_{rf}$, which can vary by an order of magnitude in the context of the high elevation mountaintop detector as shown in the middle right panel of Fig.~\ref{fig:diffacceptance}.  Conservatively assuming that the electric field can be reconstructed to within 25\% and the distance to tau decay for all events is given by the median $s_{rf}$ weighted by an $\mathcal{E}^{-2}$ spectrum, we estimate that the energy resolution -- given by the width of the $\log(\mathcal{E}_{rec}/\mathcal{E}_{\nu})$ distribution -- is 0.4 at $\mathcal{E}_{\nu}=10^{17}$~eV and rises to 0.8 at $\mathcal{E}_{\nu}=10^{20}$~eV. At $\mathcal{E}_{\nu_{\tau}} = 10^{20}$~eV, the uncertainty due to the energy transfer from the $\nu_\tau$ to the shower dominates the energy resolution, but at energies near the threshold of $\mathcal{E}_{\nu_{\tau}} = 10^{17}$~eV, the resolution may be improved through careful design of the reconstruction array or novel reconstruction techniques~\cite{Schoorlemmer:2020low}.

\section{Conclusions}\label{sec:conclusion}

A mountaintop radio beamformer is a highly efficient detector for $\nu_\tau$'s with energy greater than $10^{17}$~eV. Such a detector is complementary to existing and proposed experiments that rely on optical techniques or require ice for the interaction volume. Since the Earth-skimming technique is sensitive mainly to $\nu_\tau$'s, such an experiment can measure the single flavor neutrino spectrum at energies where flavor is challenging for other detectors to identify. This design bridges the gap in energy between the optical component of the proposed IceCube-Gen2 and radio experiments such as ARA, ARIANNA, and ANITA.

Based on the trade studies presented here, with only 100 stations using the 30-80 MHz reference design, several neutrinos could be detected in 5 years or the ultra-high energy flux could be constrained. The higher frequency design is optimized for a lower energy threshold; however, given that the differences in acceptance between the higher frequency and lower frequency design are within a factor of 2, the radio backgrounds at a given site and the desired pointing resolution and cosmic ray rejection will likely drive the frequency choice. Regardless of the frequency band chosen, a mountaintop radio detector can reach optimistic predictions of cosmogenic neutrino models assuming ultra-high energy cosmic rays are pure iron in only 3 years and with 1000 stations, noting that the possibility of iron-only primaries is disfavored by measurements of the composition of ultra-high energy cosmic rays~\cite{Auger2014_XmaxAbove178,Auger2017_Composition}.

The mountaintop radio detector design is scalable both because of its simplicity and cost effectiveness and because many sites around the world can accommodate such detector. Sites need to have mountain ridges with at least 2~km prominence with a clear view of the horizon and be reasonably radio quiet. With careful planning, full sky coverage or a deep exposure towards a particular region of the sky could be achieved through a thoughtfully designed global network of high-elevation radio stations. The expected degree scale pointing will enable searches for point sources of neutrinos and efficient background rejection of down-going cosmic rays. The performance of a prototype instrument and event reconstruction capabilities of particular design will be the subject of future work.
\\
\\
\noindent{\it Acknowledgements:} 
We thank David Saltzberg for insightful and thoughtful conversations.
S.~W. thanks the National Science Foundation for support through CAREER Award \#1752922. S.~W., C.~P., M.~V., and M.~S.-K. thank the Bill and Linda Frost Fund at the California Polytechnic State University for their support. 
Part of this work was carried out at the Jet Propulsion Laboratory, California Institute of Technology, under a contract with the National Aeronautics and Space Administration.
J.~A-M. and E.~Z. thank the financial support of Ministerio de Econom\'\i a, Industria y Competitividad 
(FPA 2017-85114-P), 
Xunta de Galicia (ED431C 2017/07) and RENATA Red Nacional Tem\'atica de Astropart\'\i culas 
(FPA2015-68783-REDT).
This work is supported by the Mar\'\i a de Maeztu Units of Excellence program MDM-2016-0692 
and the Spanish Research State Agency.
This work is co-funded by the European Regional Development Fund (ERDF/FEDER program).
W.~C. thanks grant \#2015/15735-1, S\~ao Paulo Research Foundation (FAPESP).
The UChicago group thanks the Research Corporation for Science Advancement and the Sloan Foundation, as well as NSF Award \#1607555 and NASA Award \#80NSSC18K0231. K.~H. thanks the NSF for support through the Graduate Research Fellowship Program Award DGE-1746045. 
We gratefully acknowledge support from the NSF-funded White Mountain Research Station and especially the expertise and support provided by from Jeremiah Eames and Steven Devanzo.
\clearpage

\appendix


\section{Geometric Acceptance of Detectors at Altitude}\label{app:geometric_acceptance}

\subsection{Differential Acceptance}
In this section we show that the differential acceptance of a mountaintop detector can be approximated analytically up to the evaluation of the integral over the neutrino directions convolved with their observation probabilities. We then describe the numerical estimates of the differential acceptance used in Section~\ref{sec:refdesign}.

We start by deriving the geometric component of the differential acceptance.  
Taking the differential of Equation~\ref{eq:acceptance_simple} with respect to $\theta_E$ yields
\begin{equation}
\frac{d\langle A\Omega \rangle } {d\theta_E}
= 
2\pi R_E^2 \sin\theta_E 
\iint d\Omega_{\nu} \ \hat{\mathbf{r}}_{\nu}\cdot \hat{\mathbf{n}}_E \ 
p_\mathrm{obs}(\mathcal{E}_{\nu}, \mathbf{x}_E, \mathbf{\hat{r}}_\nu) \ 
\Theta(\hat{\mathbf{r}}_{\nu}\cdot \hat{\mathbf{n}}_E).
\end{equation}
The cosine rule provides the relations between $\theta_\mathrm{emerge}$ and $\theta_E$ (defined in Figure~\ref{fig:geom_concept}), the zenith angle $\theta_{z}$ at the detector of the line of sight to the exit point, the radius of Earth $R_E$, the height of the detector $h$ and the distance between the detector and the exit point $x$ as follows
\begin{equation}
x^2 = R_E^2 + (R_E+h)^2 - 2R_E(R_E+h)\cos\theta_E
\label{eqn:lawcosines1}
\end{equation}
\begin{equation}
R_E^2 = x^2 + (R_E+h)^2 + 2x(R_E+h)\cos\theta_{z}
\label{eqn:lawcosines2}
\end{equation}
\begin{equation}
(R_E+h)^2 = R_E^2 + x^2 + 2R_E x \sin\theta_\mathrm{emerge}.
\label{eqn:lawcosines3}
\end{equation}
Note, however, that in this treatment we are approximating the emergence angle of the particle as corresponding to the emergence angle of the line of sight to the exit point. In this application, these differ approximately by the Cherenkov angle ($\sim 1^\circ$). The zenith angle $\theta_z$ corresponds to $\theta_{Z,RF}$ in the right side of Fig.~\ref{fig:geom_concept} if the decay were at the exit point and $\theta_{view}$ is 0 degrees. 
$\cos \theta_z$ is a function of $\theta_E$ and given by 
\begin{equation}
\cos\theta_z = \frac{(R_E + h)^2 - R_E^2 - x^2} {2R_E x }
\end{equation}
and $x$ is given by
\begin{equation}
x=\sqrt{(R_E + h)^2 + (R_E)^2  - 2R_E(R_E + h)\cos\theta_E}.
\label{eq:x}
\end{equation}
Differentiating Eqns.~\ref{eqn:lawcosines1}-\ref{eqn:lawcosines3} above with respect to $\cos\theta_E$ and rearranging terms yields the relations
\begin{equation}
\frac{dx}{d\cos\theta_E} = -\frac{R_E(R_E+h)}{x},
\end{equation}

\begin{equation}
\frac{d\cos\theta_{z}}{d\cos\theta_E} = \frac{R_E}{x}\left(1+\frac{R_E+h}{x}\cos\theta_{z}\right),
\end{equation}

\begin{equation}
\frac{d\sin\theta_\mathrm{emerge}}{d\cos\theta_E} = \frac{R_E+h}{x}\left(1+\frac{R_E}{x}\sin\theta_\mathrm{emerge}\right).
\end{equation}
Applying the chain rule we have

\begin{equation}
\frac{d\langle A\Omega \rangle} {d\theta_z}
= 
2\pi R_E^2 \frac{R_E}{x}\left(1+\frac{R_E+h}{x}\cos\theta_{z}\right) \sin\theta_{z} 
\iint d\Omega_{\nu} \ \hat{\mathbf{r}}_{\nu}\cdot \hat{\mathbf{n}}_E \ 
p_\mathrm{obs}(\mathcal{E}_{\nu}, \mathbf{x}_E, \mathbf{\hat{r}}_\nu) \ 
\Theta(\hat{\mathbf{r}}_{\nu}\cdot \hat{\mathbf{n}}_E)
\label{eqn:diffaccep_zenith}
\end{equation} 

and
\begin{equation}
\frac{d\langle A\Omega \rangle} {d\theta_\mathrm{emerge}}
= 
-2\pi R_E^2 \frac{R_E+h}{x}\left(1+\frac{R_E}{x}\sin\theta_\mathrm{emerge}\right)
 \cos\theta_\mathrm{emerge} 
 \iint d\Omega_{\nu} \ \hat{\mathbf{r}}_{\nu}\cdot \hat{\mathbf{n}}_E \ 
p_\mathrm{obs}(\mathcal{E}_{\nu}, \mathbf{x}_E, \mathbf{\hat{r}}_\nu) \ 
\Theta(\hat{\mathbf{r}}_{\nu}\cdot \hat{\mathbf{n}}_E).
\label{eqn:diffaccep_emerge}
\end{equation}

Evaluating the differential acceptances above can be only a matter of changing the analytically determined factors in front of the integral with respect to $\Omega_\nu$. 

The differential acceptances shown in the left panels of Fig.~\ref{fig:diffacceptance} are the result of evaluating Eqn.~\ref{eqn:diffaccep_emerge} numerically using the Monte Carlo, but were validated with the analytical prescription described above. The total integrated geometric acceptance $\langle A \Omega \rangle_g$ is computed via the Monte Carlo. The differential acceptance is estimated by summing the observation probability $p_\mathrm{obs}$ over $N(\theta_{\mathrm{emerge}})$ events in a given bin of width $\Delta \theta_{\mathrm{emerge}}$
\begin{equation}
    \frac{d\langle A\Omega \rangle} {d\theta_\mathrm{emerge}} = \frac{\langle A \Omega \rangle_g}{N} \frac{N(\theta_{\mathrm{emerge}})}{\Delta \theta_{\mathrm{emerge}} }  \sum_{i}^{N(\theta_{\mathrm{emerge}})} p_\mathrm{obs}(\theta_{\mathrm{emerge}} )
\end{equation}
Note that in this case, $\theta_\mathrm{emerge}$ corresponds to the emergence angle of the particles sampled and is not approximated to the line of sight to the exit point.


\subsection{Geometric Acceptance}
In this section we provide a derivation of the geometric acceptance, describe the numerical estimation approach used in this paper, and compare to results for high-altitude detectors from Motloch, Privitera, and Hollon~\cite{Motloch2014}. We begin with Equation~\ref{eq:geom_acceptance}
\begin{equation}
\langle A\Omega \rangle_g
= 
R_E^2 \iint d\Omega_E \iint d\Omega_{\nu} \ \hat{\mathbf{r}}_{\nu}\cdot \hat{\mathbf{n}}_E \ 
\Theta(\theta_\mathrm{cut}-\theta_d) \ 
\Theta(\hat{\mathbf{r}}_{\nu}\cdot \hat{\mathbf{n}}_E).
\end{equation}
The term $\Theta(\hat{\mathbf{r}}_{\nu}\cdot \hat{\mathbf{n}}_E)$ requires that $\hat{\mathbf{r}}_{\nu}\cdot \hat{\mathbf{n}}_E>0$ because
we are only considering neutrino trajectories exiting the Earth. The absence of this term would mean that we subtract the particles entering the reference area of integration. 

Let us handle the integration by the following choice of coordinate system. At a given exit point with Earth latitude angle $\theta_E$, let the $\hat{\mathbf{z}}$ axis lie along the line connecting the exit point to the observatory with $\hat{\mathbf{x}}$ in the plane of the figure and $\hat{\mathbf{y}}$ pointing into the page. 
In this coordinate system, the vectors are given by $\hat{\mathbf{n}}_E = \sin\theta_z\hat{\mathbf{x}}+\cos\theta_z\hat{\mathbf{z}}$ and $\hat{\mathbf{r}}_{\nu} = \sin\theta_{\nu}\cos\phi_{\nu}\hat{\mathbf{x}}+\sin\theta_{\nu}\sin\phi_{\nu}\hat{\mathbf{y}}+\cos\theta_{\nu}\hat{\mathbf{z}}$ and $\theta_d=\theta_{\nu}$.
The dot product can be expressed as 
\begin{equation}
\hat{\mathbf{r}}_{\nu}\cdot \hat{\mathbf{n}}_E = \sin\theta_z\sin\theta_{\nu}\cos\phi_{\nu} + \cos\theta_z\cos\theta_{\nu}.
\label{eq:expanded_dot_prod}
\end{equation}
 For the geometries of interest here, the conditions $\hat{\mathbf{r}}_{\nu}\cdot \hat{\mathbf{n}}_E>0$, $\sin\theta_{z}>0$, $\cos\theta_{z}>0$, $\sin\theta_{\nu}>0$ and $\cos\theta_{\nu}>0$ require that $\cos\phi_{\nu}>-\cot\theta_z \cot \theta_{\nu}$. 

In order to compare with the previous result of Ref.~\cite{Motloch2014}, it is illustrative to derive their result. In order to do that from Equation~\ref{eq:geom_acceptance}, we must ignore the $\Theta(\hat{\mathbf{r}}_{\nu}\cdot \hat{\mathbf{n}}_E)$ term in the integrand as well as the first term in the sum on the right hand side of Equation~\ref{eq:expanded_dot_prod} to arrive at
\begin{equation}
\langle A\Omega \rangle_g = 
R_E^2 \iint d\Omega_E 
\iint d\Omega_{\nu} \  \cos\theta_z\cos\theta_{\nu} \ 
\Theta(\theta_\mathrm{cut}-\theta_d). \\
\end{equation}
The integral with respect to $\Omega_E$ is limited to the horizon with $\cos\theta_{E,\mathrm{horz}}=R_E/(R_E+h)$ while the integral with respect to $\theta_\nu$ is limited to $\theta_\mathrm{cut}$ using the choice of coordinate system discussed earlier in this section. This reduces the expression to 
\begin{equation}
\langle A\Omega \rangle_g 
 = 
2\pi^2 \sin^2\theta_\mathrm{cut} R_E^2 \int^1_{\cos\theta_E, \mathrm{horz}}d(\cos\theta_E) \cos\theta_{z} \end{equation}
From Equation~\ref{eq:x}, we can substitute the variable $\cos\theta_E$ for $x$ using
\begin{equation}
\frac{d\cos\theta_E}{dx}=\frac{-x}{R_E(R_E+h)}
\end{equation}
 to obtain
\begin{equation}
\langle A\Omega \rangle_g = 
2\pi^2 \sin^2\theta_\mathrm{cut} R_E^2 \int^{\sqrt{h(2R_E+h)}}_h \frac{dx}{2R^2_E(R_E+h)} \left[(R_E+h)^2-R_E^2-x^2\right].
\end{equation}
Evaluating the integral and some algebra yields 
\begin{equation}
\langle A\Omega \rangle_g = 
2\pi^2 \sin^2\theta_\mathrm{cut} \frac{\left[h(2R_E+h)\right]^{3/2}-h^2\left(3R_E+h\right)}{3R_E(R_E+h)}.
\label{eq:MHP14_geom_acc}
\end{equation}

In Table~\ref{tbl:geom_acc_comparison}, we compare the values resulting from numerical integration of Equation~\ref{eq:geom_acceptance} and the results of MHP14 (Equation~\ref{eq:MHP14_geom_acc}). The discrepancies are large for lower elevations and narrower $\theta_d$  cuts ($\theta_\mathrm{cut}$) while agreement is at the 1\% level for high altitudes and narrow $\theta_d$, which was the focus of MHP14.

\begin{table}
\begin{center}
  \begin{tabular}{ | c | c | c | c| c |}
    \hline
          &            & $\langle A\Omega \rangle_g$, km$^2$sr  & $\langle A\Omega \rangle_g$, km$^2$sr  & \\ 
    h, km & $\theta_\mathrm{cut}$ &  This work                           & MHP14 & Difference \\ \hline\hline
    1 &   $3^\circ$ & 6.6 & 4.0   & 65\%  \\ 
    3 &   $3^\circ$ & 26.9 & 20.7 & 30\%  \\
    37 &  $3^\circ$ & 880 & 841   & 4.6\% \\
    600 & $3^\circ$ & 40,618 & 40,209 & 1.0\% \\  \hline
    1 &   $1^\circ$ & 0.51 & 0.45     & 13\% \\ 
    3 &   $1^\circ$ & 2.4 & 2.3       & 4.4\% \\
    37 &  $1^\circ$ & 94.5 & 93.5     & 1.1\% \\
    600 & $1^\circ$ & 4,499 & 4,471   & 0.6\% \\
    \hline
  \end{tabular}
  \end{center}
 \caption{Comparison of the difference in geometric acceptance calculated in this work and in Ref.~\cite{Motloch2014} (MHP14) for mountaintop (1-3~km), balloon-borne (37~km), and satellite (600~km) detector heights $h$ and varying detectable view angles $\theta_\mathrm{cut}$. See Fig.~\ref{fig:geom_concept} and the text for definitions of the geometric parameters.}
\label{tbl:geom_acc_comparison}
\end{table}

\begin{figure}[!ht]
{\includegraphics[width=0.49\textwidth]{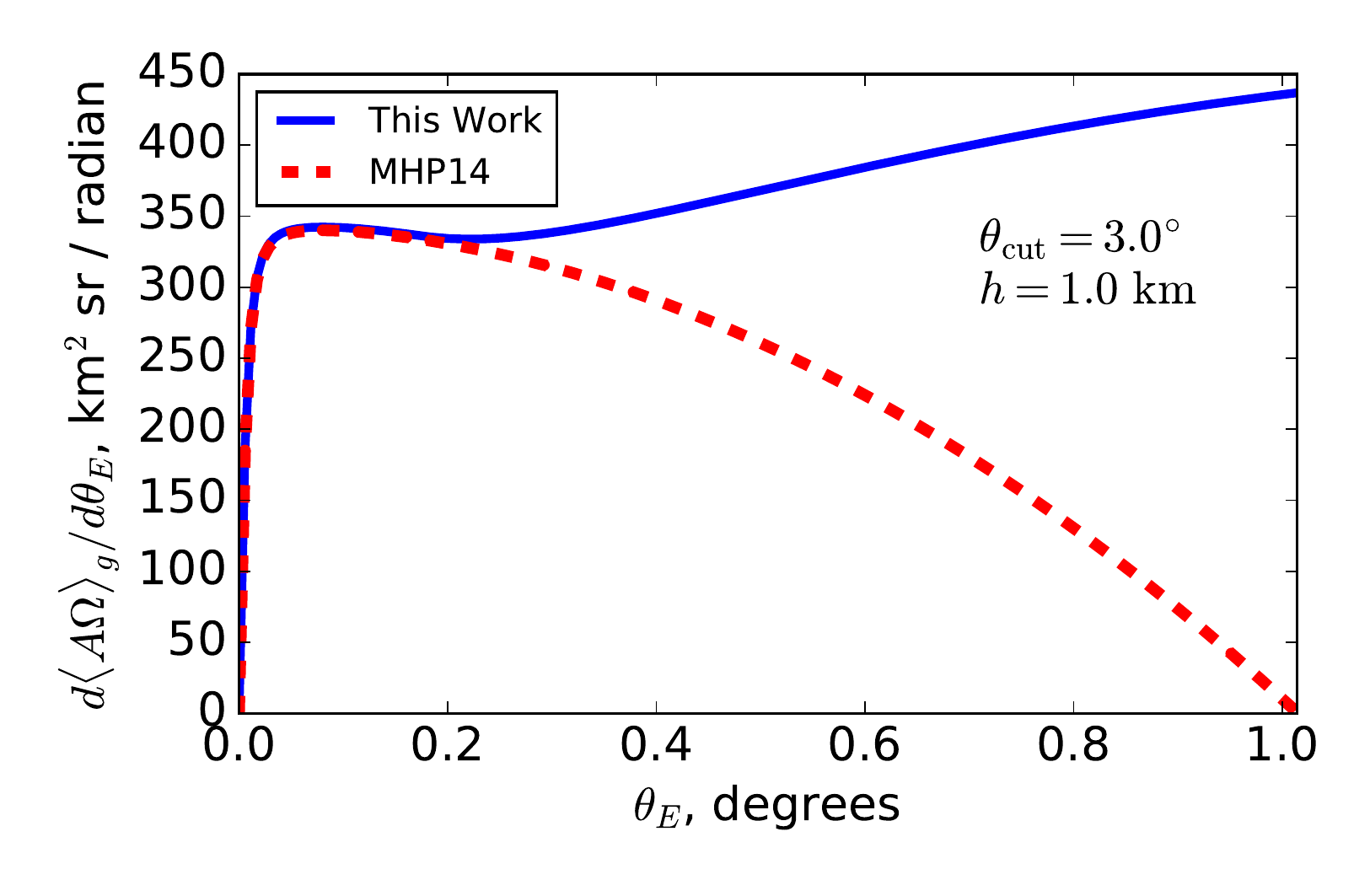}}
{\includegraphics[width=0.49\textwidth]{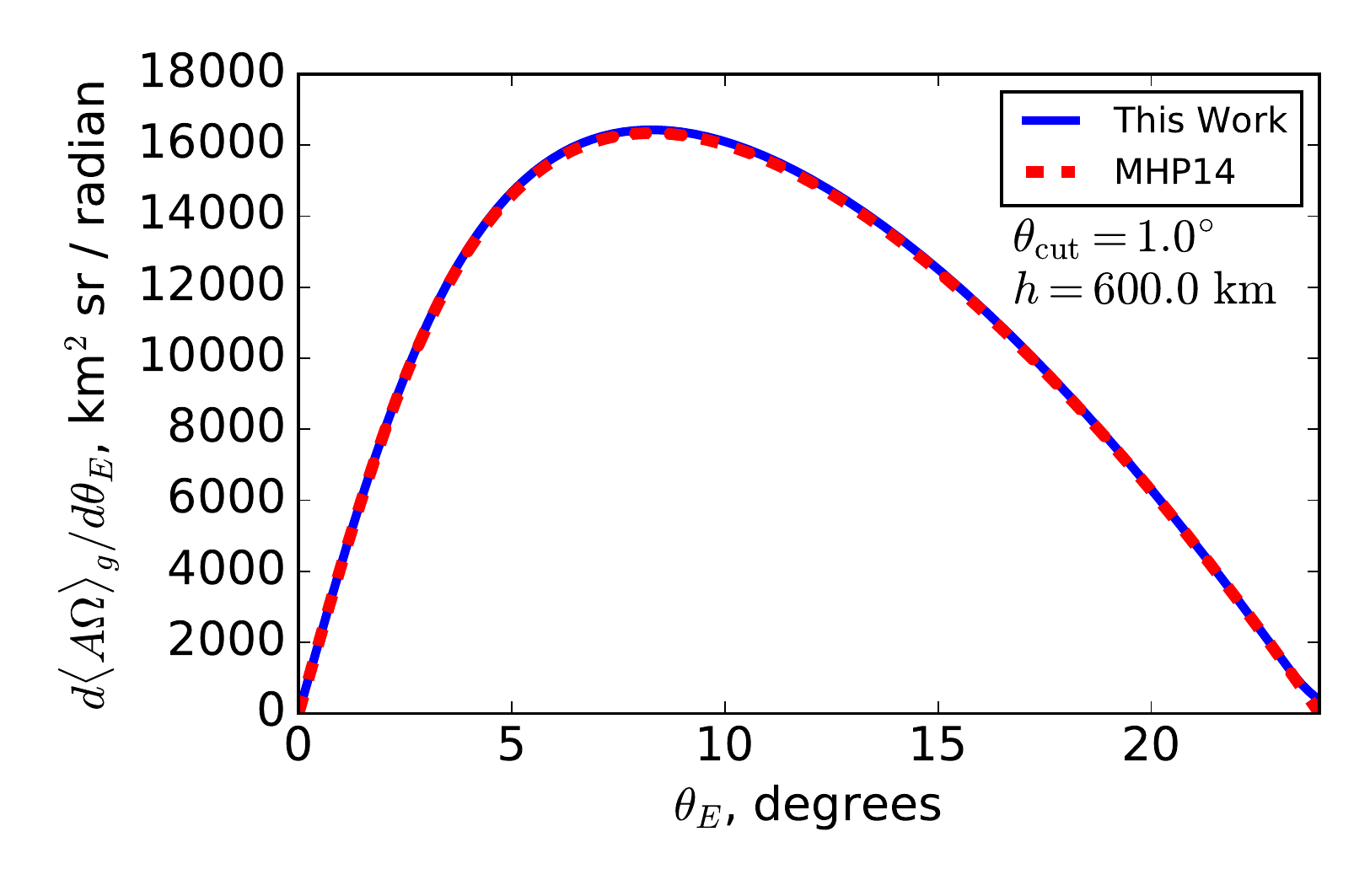}}
\caption{Comparison of the geometric acceptance estimates of MHP14 and this work. (Left) There is a significant difference in the differential geometric acceptance with respect to Earth angle $\theta_E$ going towards the horizon for cut angle $\theta_{cut}=3.0^\circ$ on $\theta_d$ at elevation $h=1.0$~km. This can be traced to MHP14 ignoring the first term in the sum in square brackets in the integrand of Equation~\ref{eq:geom_diff_acc_integral}. (Right) Differential geometric acceptance with respect to Earth angle for $\theta_\mathrm{cut}=1.0^\circ$ and $h=600$~km altitude. In this case, the approach of MHP14 (which was developed for this case) and this work are in significantly better agreement. Despite the small difference, note that MHP14 always systematically underestimates the geometric acceptance because their approach removes particles that enter and exit the surface area of the Earth visible to the detector.  }
\label{fig:geom_diff_acc}
\end{figure}

The interpretation of this discrepancy is most easily understood by comparing the differential geometric acceptance
\begin{equation}
\frac{d\langle A\Omega \rangle_g}{d\theta_E}
= 
(2\pi R_E^2 \sin\theta_E )
\Bigg(\iint d\Omega_{\nu} \ \left[\sin\theta_z\sin\theta_{\nu}\cos\phi_{\nu} + \cos\theta_z\cos\theta_{\nu}\right]\ 
\Theta(\theta_\mathrm{cut}-\theta_d) \ 
\Theta(\hat{\mathbf{r}}_{\nu}\cdot \hat{\mathbf{n}}_E)\Bigg).
\label{eq:geom_diff_acc_integral}
\end{equation}
with the version using the approximations needed to derive the MHP14 result
\begin{equation}
\frac{d\langle A\Omega \rangle_g }{d\theta_E}
= 
\left(2\pi R_E^2 \sin\theta_E\right) \left(\pi \sin^2\theta_\mathrm{cut} \cos\theta_{z}\right),
\end{equation}
where we have put the terms in parenthesis to highlight the differences between the two approaches. The differential acceptance curves are shown for several different scenarios in Figure~\ref{fig:geom_diff_acc}. It is clear that the approximations used for MHP14 tend to neglect the contributions from the horizon. This approximation is valid for small values of $\theta_\mathrm{cut}$, which is a reasonable approximation for detectors at stratospheric balloon or spacecraft altitudes. However, at mountain elevation scales and observing at low frequencies, this result tends to underestimate the acceptance, particularly near the horizon. 

\clearpage

\bibliographystyle{JHEP}
\bibliography{references}

\end{document}